\documentclass[aps,prb,reprint,
citeautoscript,superscriptaddress,showpacs]{revtex4-1}

\usepackage{graphicx,epsfig,float}
\usepackage{amssymb}
\usepackage{amsmath}
\usepackage[varg]{txfonts}
\allowdisplaybreaks
\begin{document}
\title{Structure and energetics of carbon, hexagonal boron nitride and carbon/hexagonal boron nitride single-layer and bilayer nanoscrolls}

\author{Andrei I. Siahlo}
\affiliation{Physics Department, Belarusian State University, Nezavisimosti Ave. 4, Minsk 220030, Belarus}

\author{Nikolai A. Poklonski}
\email{poklonski@bsu.by}
\affiliation{Physics Department, Belarusian State University, Nezavisimosti Ave. 4, Minsk 220030, Belarus}

\author{Alexander V. Lebedev}
\email{allexandrleb@gmail.com}
\affiliation{Kintech Lab Ltd., 3rd Khoroshevskaya Street 12, Moscow 123298, Russia}

\author{Irina V. Lebedeva}
\email{liv\_ira@hotmail.com}
\affiliation{Nano-Bio Spectroscopy Group and ETSF, Universidad del Pa\'is Vasco, CFM CSIC-UPV/EHU, 20018 San Sebastian, Spain}

\author{Andrey M. Popov}
\email{popov-isan@mail.ru}
\affiliation{Institute for Spectroscopy of Russian Academy of Sciences, Troitsk, Moscow 108840, Russia}

\author{Sergey A. Vyrko}
\affiliation{Physics Department, Belarusian State University, Nezavisimosti Ave. 4, Minsk 220030, Belarus}

\author{Andrey A. Knizhnik}
\email{kniznik@kintechlab.com}
\affiliation{Kintech Lab Ltd., 3rd Khoroshevskaya Street 12, Moscow 123298, Russia}

\author{Yurii E. Lozovik}
\affiliation{Institute for Spectroscopy of Russian Academy of Sciences, Troitsk, Moscow 108840, Russia}
\affiliation{N.L.\,Dukhov All-Russia Research Institute of Automatics, 127055 Moscow, Russia}
\affiliation{Moscow Institute of Electronics and Mathematics, National Research University Higher School of Economics, 101000 Moscow, Russia}

\begin{abstract}
Single-layer and bilayer carbon and hexagonal boron nitride nanoscrolls as well as nanoscrolls made of bilayer graphene/hexagonal boron nitride heterostructure are considered.
Structures of stable states of the corresponding nanoscrolls prepared by rolling single-layer and bilayer rectangular nanoribbons are obtained based on the analytical model and numerical calculations.
The lengths of nanoribbons for which stable and energetically favorable nanoscrolls are possible are determined. Barriers to rolling of single-layer and bilayer nanoribbons into nanoscrolls and barriers to nanoscroll unrolling are calculated. Based on the calculated barriers nanoscroll lifetimes in the stable state are estimated. Elastic constants for bending of graphene and hexagonal boron nitride layers used in the model are found by density functional theory calculations.
\end{abstract}

\maketitle

\section{Introduction}

A set of methods for synthesis of carbon nanoscrolls (CNSs) have been elaborated in the recent decade including electrochemical exfoliation of graphite leading to dispersion of monolayer graphene sheets \cite{Viculis2003, Shioyama2003, Savoskin2007}, high energy ball milling of graphite \cite{Li2005}, chemical vapor deposition \cite{Ruland2003, Chuvilin2009, Schaper2011}, electrostatic deposition of graphene sheets in hydrogen atmosphere \cite{Sidorov2009}, microexplosion method \cite{Zeng2011,Zeng2012}, use of microwave sparks in liquid nitrogen \cite{Zheng2011}, rolling of a graphene layer on a substrate immersed into isopropyl alcohol solution \cite{Xie2009} and rolling around water nanodroplet \cite{Mirsaidov2013} and around nanowires \cite{Yan2013}. A CNS-based foam with a high specific capacitance and a very low density has been produced \cite{Zheng2016}. A CNS-based nanoelectronic device has been also fabricated \cite{Xie2009}. These advances in CNS synthesis generate considerable interest to electronic \cite{Pan2005, Rurali2006, Chen2007, Dong2016}, optical \cite{Pan2005}, electric \cite{Schaper2011, Zeng2011, Zeng2012, Zheng2011} and mechanical \cite{Schaper2011} properties and possible applications of CNSs. Namely, CNSs hold much promise for applications in supercapacitors \cite{Zeng2012, Yan2013, Zheng2016}, batteries \cite{Yan2013}, chemical sensors \cite{Karimi2014, Khaledian2015}, nanofluidic devices \cite{Shi2010, Mirsaidov2013}, nanoelectromechanical systems \cite{Rurali2006} and for hydrogen storage \cite{Mpourmpakis2007, Coluci2007, Braga2007}.

The following theoretical approaches have been used to consider structure and energetics of nanoscrolls, which hold the key to understanding their properties, application and formation mechanisms. First, the structure and energetics of CNSs have been studied by density functional theory (DFT) calculations in the local density approximation (LDA \cite{Ceperley1980}) \cite{Chen2007}. However, such calculations considerably underestimate the interlayer interaction energy of graphite \cite{Lebedeva2010, Kolmogorov2005, Wang2014, Reguzzoni2012, Lebedeva2011a}. Furthermore, DFT calculations do not allow to consider systems consisting of more than several hundred atoms. Second, a set of semiempirical atomistic approaches has been used. Structure of SCNSs was studied using a chain model in which a row of carbon atoms parallel to the nanoscroll axis is considered as one particle \cite{Savin2015}. The processes of SCNS \cite{Shi2010a} and SBNNS \cite{Perim2009} rolling have been simulated by molecular dynamics. To use successfully such semiempirical atomistic approaches tedious work on elaboration of appropriate interatomic potentials is necessary. However, the parameters of classical interatomic potentials for carbon have not been fitted to reproduce bending elastic energies of graphene or boron nitride layers. For example, the bending elastic constant for a graphene layer calculated used different versions of the popular Brenner potential is about 2 times smaller \cite{Arroyo2004} than the constant obtained by previous \emph{ab initio} calculations \cite{Kudin2001, Lebedeva2012, Gulseren2002, Cherian2007, Sanchez1999, Kurti1998} and by DFT calculations performed in the present paper. Whereas semiempirical atomistic approaches allow to consider nanoscrolls consisting of thousands of atoms, very few examples of a large size can be considered, which is not sufficient to study the structure and energetics as functions of nanoscroll dimensions. Third, two types of analytical and semianalytical models have been elaborated \cite{Fogler2010, Shi2010a, Shi2011, Siahlo2017, Yamaletdinov2017}. These models allow to calculate dependences of structural and energetic characteristics of a nanoscroll on its dimensions using values of bending elastic constants and interlayer interaction energies obtained in the experiment or by high-level {\it ab initio} methods. The first of the models is based on an approximate consideration of the potential energy of a single-layer nanoscroll with a large number of layers and gives an expression for dimensions of such nanoscroll in the stable state \cite{Shi2010a, Shi2011}. The second one gives the expression of nanoscroll potential energy \cite{Fogler2010, Siahlo2017, Yamaletdinov2017}, whereas the dimensions of nanoscroll in a stable state can be obtained by numerical calculations. Such a model allows to consider not only large nanoscrolls but also nanoscrolls with a minimal number of layers, which are promising, for example, for enhancement of adsorption properties of nanoscroll-based materials or decreasing the size of nanoscroll-based NEMS.

In addition to CNSs, graphene oxide nanoscrolls not only with a hollow cavity inside \cite{Gao2010} but also wrapped around carbon nanotubes \cite{Kim2010} and metal nanoparticles \cite{Wang2012} as well as boron nitride nanoscrolls have been obtained \cite{Chen2012, Li2013, Hwang2014}. Advances in synthesis of other 2D materials, such as hexagonal boron nitride and graphene/hexagonal boron nitride heterostructures (see Refs.~\onlinecite{Wang2017,Li2016} for recent reviews), and a variety of methods of CNS synthesis listed above allow to expect that other types of nanoscrolls can be produced in the nearest future. Such nanoscrolls can have unique properties, which are interesting for fundamental science and perspective for applications. For example, metamaterials with a negative magnetic permeability based on conductors rolled into scrolls have been considered \cite{Ramakrishna2005}. We believe that bilayer carbon/boron nitride nanoscrolls (CBNNSs) with alternating dielectric and conducting layers are promising for elaboration of such new metamaterials with negative refractive indexes and magnetic permeability. However, up to now only properties of single-layer carbon nanoscrolls \cite{Chen2007, Fogler2010, Shi2010a, Shi2011, Savin2015, Siahlo2017, Yamaletdinov2017} and rolling of single-layer boron nitride nanoscrolls \cite{Perim2009} have been considered. 

Here we use our recent semianalytical model \cite{Siahlo2017} to compare the structure and energetics of single-layer and bilayer carbon nanoscrolls (SCNSs and BCNSs) and single-layer and bilayer boron nitride nanoscrolls (SBNNSs and BBNNSs) as well as CBNNSs made of bilayer graphene/hexagonal boron nitride heterostructure. Not only characteristics of the stable state of nanoscrolls but also the barriers to rolling and unrolling are studied as functions of nanoscroll dimensions. The calculated barriers allow also to estimate the lifetimes of the nanoscrolls relative to unrolling using the Arrhenius formula and thus to determine dimensions of the smallest nanoscrolls which are sufficiently stable to be used in nanoelectromechanical systems, electronic devices and composite materials. Moreover, recent experimental and theoretical data are carefully analyzed to choose appropriate values of interlayer interaction energies used in the numerical calculations, whereas values of the bending elastic constants for graphene and hexagonal boron nitride are revised by calculations within the DFT approach. This allows us to perform the quantitative comparison of characteristics of the nanoscrolls.

The paper is organized in the following way. In Section II the choice of interlayer interaction energy values is discussed and elastic constants for bending of graphene and boron nitride layers used in the model are calculated. Section III presents the model of nanoscrolls and calculations of their structure and energetics. Our conclusions are summarized in Section IV.

\section{Interlayer interaction and elastic energy}

The present section is devoted to the choice of values for interlayer interaction energies in graphene-graphene, boron nitride-boron nitride and graphene-boron nitride systems and elastic energies of rolled graphene and boron nitride layers. Since standard DFT methods fail to describe properly weak van der Waals interaction of graphene and boron nitride layers, the values for the interlayer interaction energy are chosen on the basis of the literature review involving available experimental data and results of calculations within the random phase approximation (RPA) and quantum Monte Carlo (QMC) approach (subsection A). The values of the bending elastic constants are accessible through DFT and these calculations are performed in subsection B.

\subsection{Interlayer interaction energy}

The interlayer interaction in graphite has been studied in diverse experiments and the following data for the interlayer interaction energy were reported $-52\pm5$~meV$/$atom \cite{Zacharia2004}, $-43\pm5$~meV$/$atom \cite{Girifalco1956}, $-35\,^{+15}_{-10}$~meV$/$atom \cite{Benedict1998}, and $-31\pm2$~meV$/$atom \cite{Liu2012}. The RPA and QMC calculations of the interlayer interaction energy in graphite and graphene bilayer (for bilayers the energy is expressed in meV per atom of the upper (adsorbed) layer) in the ground-state AB stacking gave very similar values of $-48$~meV$/$atom \cite{Lebegue2010} (RPA), $-56 \pm 6$~meV$/$atom \cite{Spanu2009} (QMC), and $-35.6 \pm 1.6$~meV$/$atom \cite{Mostaani2015} (QMC), with the only exception of $-91.35$~meV$/$atom \cite{Zhou2015} (RPA). Therefore, we use as an estimate for the interlayer interaction energy of commensurate graphene layers the average of the experimental data, which corresponds to about $-40$~meV$/$atom.

The adjacent layers of nanoscrolls have, nevertheless, different curvature radii and should be considered as incommensurate. The interlayer interaction energy in such incommensurate structures can be estimated as the average of interlayer interaction energies for commensurate structures with different relative in-plane displacement of the layers \cite{Lebedeva2010, Lebedeva2011, Sachs2011, Popov2012, Lebedev2016}. Based on the potential energy surface for interaction of graphene layers fitted to the experimental data on the shear mode frequencies in bilayer, few-layer graphene and graphite, the energy cost for transition from the ground-state structure to the incommensurate one was estimated to be about 5~meV$/$atom \cite{Popov2012}. The same value can be deduced from the DFT calculations at the experimental interlayer distance \cite{Lebedeva2017} using the exchange-correlation functional of Perdew, Burke and Ernzerhof (PBE) \cite{Perdew1996} as well as PBE-D2, PBE-D3, PBE-D3(BJ), vdW-DF2 and optPBE-vdW functionals. This approach has been verified against the experimental data on the shear mode frequencies, shear modulus and the results of RPA, QMC and local second-order M{\o}ller-Plesset perturbation theory (LMP2) calculations for bilayer graphene, graphite, bilayer and bulk boron nitride \cite{Lebedeva2017}. It can be also mentioned that the LDA calculations at the optimized interlayer distance gave the result which is only 1~meV$/$atom smaller \cite{Lebedeva2010, Lebedeva2011a}. Therefore, it is reasonable to assume that the interlayer interaction energy of incommensurate graphene layers is about $-35$~meV$/$atom.

For the graphene/boron nitride heterostructure, the cost for the transition from the commensurate state with the AB1 stacking characterized by the minimal interlayer interaction energy to incommensurate states was estimated from the RPA calculations \cite{Sachs2011} to be about 7~meV$/$atom. The close value of 7.4~meV$/$atom was obtained from the vdW-DF2 calculations at the experimental interlayer distance \cite{Lebedev2017}. Using the minimal interlayer interaction energy of the commensurate heterostructure of about $-42$~meV$/$atom found from the RPA calculations, the interlayer interaction energy in the incommensurate state results to be about $-35$~meV$/$atom \cite{Sachs2011} again, the same as the selected value for pure graphene.

As long as we are aware there are no experimental data on the interlayer interaction in hexagonal boron nitride. The only reported RPA value of $-37.6$~meV$/$atom for the commensurate bilayer \cite{Zhou2015} needs further verification since the value of $-91.35$~meV$/$atom obtained for graphene in the same paper is clearly too large in magnitude. The result of the paper for the commensurate graphene/hexagonal boron nitride heterostructure of $-57.9$~meV$/$atom is also rather different from other publication \cite{Sachs2011} and from the results of the same paper for pure graphene and boron nitride bilayers, which seems somewhat surprising provided the apparent similarity in these van der Waals-bonded layered materials. Therefore, we do not have reliable data on the interaction energy of commensurate boron nitride layers. Nevertheless, it can be expected this energy is close to the interlayer interaction energies in graphene bilayer and graphene/boron nitride heterostructure. The energy cost for transition of boron nitride bilayer from the ground-state commensurate structure to the incommensurate one was estimated to be 6.3~meV$/$atom on the basis of vdW-DF2 calculations at the experimental interlayer distance \cite{Lebedeva2016}. Similar values of 6--7~meV$/$atom can be deduced from other DFT calculations at the experimental interlayer distance \cite{Lebedeva2017} and LMP2 calculations \cite{Constantinescu2013}. These data are of the same order of magnitude as those for graphene and graphene/boron nitride heterostructure and it can be roughly assumed that the interlayer interaction energy of incommensurate boron nitride layers is also $-35$~meV$/$atom.

\subsection{Bending elastic constant}

The results of previous calculations related to bending elastic energies of graphene \cite{Kudin2001, Lebedeva2012, Gulseren2002, Cherian2007, Sanchez1999, Kurti1998, Hernandez1999, Hernandez1998, Blase1994} and boron nitride \cite{Kudin2001, Kurti1998, Hernandez1999, Hernandez1998, Blase1994, Xiang2003, Baumeier2007} layers show significant scatter.
To get insight into these characteristics we have performed DFT calculations of elastic energies of carbon and boron nitride nanotubes using the VASP code \cite{Kresse1996}. The dependence of the elastic energy on the nanotube chirality is usually negligibly small \cite{Xiang2003, Sanchez1999, Kudin2001, Lebedeva2012, Hernandez1999, Hernandez1998, Baumeier2007, Blase1994}. Therefore, only armchair nanotubes are studied here. Radii of the considered nanotubes are within 10~\AA{} and the calculations are done for the minimal unit cells of the nanotubes under periodic boundary conditions. The dimensions of the rectangular model cell perpendicular to the nanotube axis are 40~\AA. The PBE functional \cite{Perdew1996} is used and the projector augmented-wave method (PAW) \cite{Kresse1999} is applied to describe the interaction of valence electrons with the core. The integration over the Brillouin zone is carried out according to the Monkhorst--Pack scheme \cite{Monkhorst1976} with 36 $k$-points along the nanotube axis. The maximum kinetic energy of plane waves is 550~eV. The convergence threshold of the self-consistent field is $10^{-8}$~eV. The size of the model cell along the nanotube axis and positions of atoms within the model cell are optimized so that the residual forces are within 0.01~eV/\AA. The calculations for flat graphene and boron nitride layers have been also performed to extract the elastic energies of the nanotubes. In these calculations the rectangular model cells comprising 4 atoms are considered with the 20~\AA{} vacuum gap between the periodic images of the layers. The 36$\times$40$\times$1 k-point grid is used, where the first number corresponds to the armchair direction and second one to the zigzag direction. The relative energy of nanotubes per atom is computed as $E_\text{rel} = E_\text{NT}-E_\text{FL}$, where $E_\text{NT}$ and $E_\text{FL}$ are the energies per atom in the nanotube and flat layer, respectively.

\begin{figure}
\centering
\includegraphics{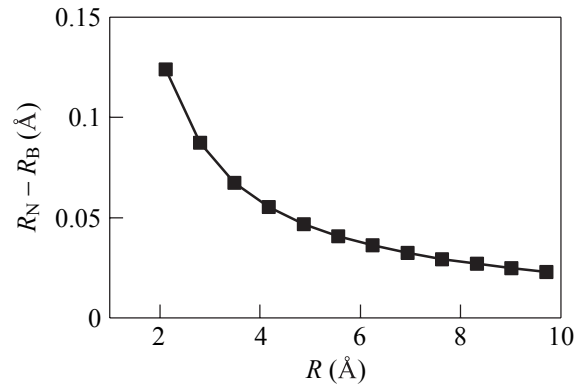}
\caption{Calculated difference $R_\text{N}- R_\text{B}$ (in \AA) in the radii of the cylinders formed by nitrogen and boron atoms in geometrically optimized boron nitride nanotubes as a function of the average nanotube radius $R$ (in \AA).}\label{fig:rad_diff}
\end{figure}
The same as in previous publications \cite{Blase1994, Kudin2001, Xiang2003, Hernandez1999, Hernandez1998, Baumeier2007}, geometry optimization performed in the present paper leads to displacement of boron and nitrogen atoms inward and outward, respectively, of the initially smooth wall. As a result, two cylinders are formed, where the inner one is composed of boron and the outer one of nitrogen. The difference in the radii of the cylinders formed by nitrogen, $R_\text{N}$, and boron, $R_\text{B}$, decreases with increasing the average radius $R$ (Fig.~\ref{fig:rad_diff}). This average radius is used below to study the dependence of the elastic energy on the curvature radius.

Though we subtract the energies per atom in the flat layers we should admit that the calculations for nanotubes and flat layers are not fully consistent, e.g., the model cells are different. This small inconsistency, however, is not important for calculations of the elastic energy as long as we assume that the calculated relative energy $E_\text{rel} = E_\text{el} + E_0$ is given by the sum of the elastic energy $E_\text{el}$ and small correction $E_0$ to the energy of the flat layer per atom. We also assume that the elastic energy depends on the average curvature radius $R$ as $E_\text{el} = C/R^p$. The parameters corresponding to the best fit to this dependence are listed in Table \ref{table:approx}. According to our results, both for carbon and boron nitride nanotubes the deviation from the classical quadratic law with $p=2$ is negligible.

\begin{table*}
\caption{Parameters of approximation of the energy of nanotubes relative to the graphene layer obtained by DFT calculations according to the expression $E_\text{rel} = E_0 + C/R^p$.}
\renewcommand{\arraystretch}{1.5}
\resizebox{0.6\textwidth}{!}{%
\begin{tabular}{*{6}{c}}\hline
Method & Nanotubes & $E_0$ (meV$/$atom) & $C$ (eV\,\AA$^2/$atom) & $p$ & Ref. \\\hline
& & & Carbon & & \\\hline
PBE & armchair & $9.59 \pm 0.13$ & $2.146 \pm 0.013$ & $2.053 \pm 0.005$ & This work \\\hline
LDA & armchair & $-18 \pm 7$ & $1.1 \pm 0.2$ & $1.5 \pm 0.2$ & [\onlinecite{Kurti1998}] \\\hline
LDA & zigzag & $2 \pm 2$ & $2.07 \pm 0.14$ & $2.06 \pm 0.07$ & [\onlinecite{Kurti1998}] \\\hline
& & & Boron nitride & & \\\hline
PBE & armchair & $6.21 \pm 0.05$ & $1.3227 \pm 0.0010$ & $1.9943 \pm 0.0010$ & This work \\\hline
LDA & armchair & $26.5 \pm 0.8$ & $0.83 \pm 0.02$ & $1.75 \pm 0.02$ & [\onlinecite{Baumeier2007}] \\\hline
LDA & zigzag & $53 \pm 7$ & $1.56 \pm 0.09$ & $2.53 \pm 0.12$ & [\onlinecite{Baumeier2007}] \\\hline
\end{tabular}%
}
\label{table:approx}
\end{table*}

Let us discuss the reasons of the deviation of exponent $p$ from 2 based on our calculations and literature data.
The cases of deviation of exponent $p$ from 2 for boron nitride nanotubes have been reported in literature. In Ref.~\onlinecite{Kudin2001} approximation $E_\text{rel} = 1.60/R^{1.94} - 1.65 /R^{1.95}$ (PBE) was obtained for armchair boron nitride nanotubes and in Ref.~\onlinecite{Xiang2003} the best fit by the power law gave $E_\text{rel} = 1.4497/R^{2.09481}$ (LDA) both for armchair and zigzag boron nitride nanotubes. Some deviation from $p=2$ is also observed when we fit the \textit{ab initio} data presented in Refs.~\onlinecite{Baumeier2007} and \onlinecite{Kurti1998} (Table \ref{table:approx}). On the other hand, very small deviations from $p = 2$ were found by tight-binding calculations in Ref.~\onlinecite{Hernandez1999}, namely $E_\text{rel} = 2.0/R^{2.083}$ and $E_\text{rel} = 2.2/R^{1.996}$ for armchair and zigzag carbon nanotubes, respectively, and $E_\text{rel} = 1.4/R^{1.984}$ and $E_\text{rel} = 1.4/R^{1.980}$ for armchair and zigzag boron nitride nanotubes, respectively. The scatter in the exponent $p$ in different papers indicates that its deviation from $p = 2$ is of computational rather than of fundamental nature. In particular, if we ignore the energy offset $E_0$, which is within the typical error of DFT calculations, in approximation of our data, $p$ gets much farther from 2, to 1.77 for carbon nanotubes and 1.87 for boron nitride nanotubes. Increasing the number of considered nanotubes of different radii also improves the agreement with the classical law with $p = 2$ according to our calculations.

\begin{table}
\caption{Parameters of approximation of the energy of nanotubes relative to the graphene layer obtained by DFT calculations according to the expression $E_\text{rel} = E_0 + C/R^2$ ($p=2$).}
\renewcommand{\arraystretch}{1.3}
\resizebox{0.5\textwidth}{!}{%
\begin{tabular}{*{5}{c}}
\hline
Method & Nanotubes & $E_0$ (meV$/$atom) & $C$ (eV\,\AA$^2/$atom) & Ref. \\\hline
& & Carbon & & \\\hline
PBE & armchair & $8.1 \pm 0.3$ & $2.010 \pm 0.007$ & This work \\\hline
PBE & armchair/zigzag & & 1.95 & [\onlinecite{Kudin2001}] \\\hline
PBE & armchair & & 1.92--1.96 & [\onlinecite{Lebedeva2012}] \\\hline
PBE & zigzag & & 1.93--1.98 & [\onlinecite{Lebedeva2012}] \\\hline
PW91 & armchair & & 2.02--2.17 & [\onlinecite{Cherian2007}] \\\hline
PW91 & zigzag & & 2.14 & [\onlinecite{Gulseren2002}] \\\hline
LDA & armchair & $-2.2 \pm 1.0$ & $2.09 \pm 0.04$ & [\onlinecite{Kurti1998}] \\\hline
LDA & zigzag & $-0.2 \pm 0.5$ & $1.949 \pm 0.009$ & [\onlinecite{Kurti1998}] \\\hline
LDA & armchair & & 2.00 & [\onlinecite{Sanchez1999}] \\\hline
LDA & (8,4) & & 2.15 & [\onlinecite{Sanchez1999}] \\\hline
LDA & (10,0) & & 2.16 & [\onlinecite{Sanchez1999}] \\\hline
& & Boron nitride & & \\\hline
PBE & armchair & $6.43 \pm 0.04$ & $1.3280 \pm 0.0004$ & This work \\\hline
LDA & armchair & $33.8 \pm 1.1$ & $1.03 \pm 0.02$ & [\onlinecite{Baumeier2007}] \\\hline
LDA & zigzag & $18 \pm 8$ & $1.25 \pm 0.04$ & [\onlinecite{Baumeier2007}] \\\hline
\end{tabular}%
}
\label{table:approx1}
\end{table}

\begin{figure}
\centering
\includegraphics{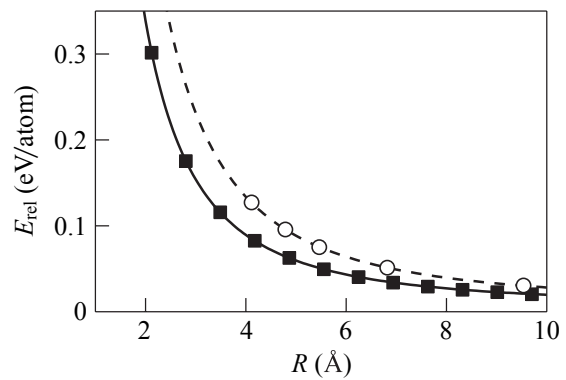}
\caption{Calculated energy $E_\text{rel}$ (in~eV$/$atom) of armchair nanotubes relative to the flat layer per atom as a function of average nanotube radius $R$ (in \AA). The results of DFT calculations for carbon (open circles) and boron nitride (filled squares) nanotubes are approximated by the dependence $E_\text{rel} = C/R^2 + E_0$, where $C=2.010$~eV\,\AA$^2/$atom and $E_0=8.1$~meV$/$atom for carbon (dashed line) and $C=1.3280$~eV\,\AA$^2/$atom and $E_0=6.43$~meV$/$atom for boron nitride (solid line).}\label{fig:elastic_energy}
\end{figure}

Therefore, it is reasonable to assume that the classical law with $p = 2$ is valid for rolling both graphene and boron nitride layers. The parameters of approximation of our data with the corresponding expression are given in Table \ref{table:approx1}. It is seen from Fig.~\ref{fig:elastic_energy} that the classical law describes our DFT data very well. The calculated bending constants are in excellent agreement with the previous \textit{ab initio} calculations (Table \ref{table:approx1}) using the LDA, PBE, and Perdew--Wang 91 (PW91) functionals \cite{Perdew92}. It should be also mentioned that the value $C=1.79$~eV\,\AA$^2/$atom close to our result for carbon nanotubes was reported previously for sulfur-terminated zigzag graphene nanoribbons \cite{Lebedeva2012} (PBE). 

The obtained coefficient $C$ for boron nitride is by a factor of one and a half smaller than that for graphene (Fig.~\ref{fig:elastic_energy}, Table \ref{table:approx1}), in agreement with the results reported previously \cite{Blase1994, Kudin2001, Hernandez1999, Hernandez1998}. This difference can be explained by the fact that the Young modulus for boron nitride is 10--20\% smaller compared to graphene \cite{Lebedeva2016, Sachs2011} as well as by buckling of boron nitride nanotubes \cite{Blase1994, Kudin2001, Xiang2003, Hernandez1999, Hernandez1998, Baumeier2007}.

\section{Structure and energetics of nanoscrolls}

\begin{figure}
\centering
\includegraphics{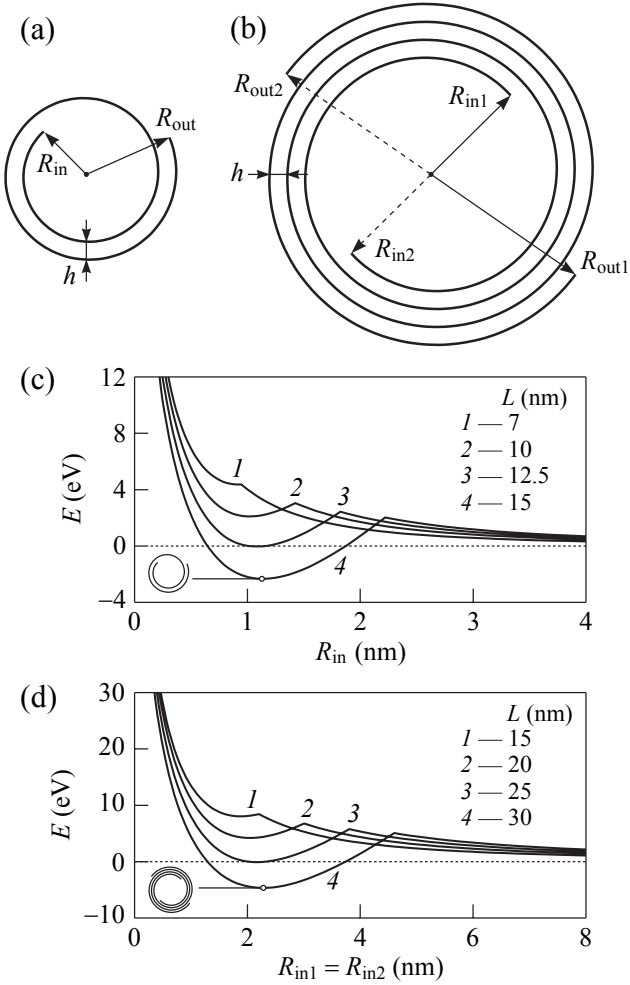}
\caption{(a), (b) Schemes of nanoscrolls with the axis perpendicular to the figure plane: (a) a single-layer nanoscroll, (b) a bilayer nanoscroll of identical layers. (c), (d) Calculated potential energy $E$ of carbon nanoscrolls per unit width (1~nm) of the initial nanoribbon as a function of the inner radius $R_\text{in}$: (c) single-layer nanoscrolls made from graphene nanoribbons with the length $L$ = 7, 10, 12.5, and 15~nm, (d)~bilayer nanoscrolls made from bilayer graphene nanoribbons with the length $L$ = 15, 20, 25, and 30~nm. The energies of bilayer nanoscrolls are measured relative to the energy of the flat bilayer graphene nanoribbon.}\label{fig:3}
\end{figure}

We consider the structure and energetics of nanoscrolls made from flat rectangular single-layer and bilayer nanoribbons with the length $L$ and width $w$. According to the analytical model of single-layer nanoscrolls \cite{Siahlo2017, Yamaletdinov2017} used here, the distance $R$ between the layer and the nanoscroll
axis is described in polar coordinates $(R,\varphi)$ by the equation of the Archimedean spiral
$R =h \varphi /2\pi$, where $h$ is the approximate distance between adjacent spiral turns
(equal to the interlayer spacing of the nanoscroll). The length $L$ is measured along the spiral line and the width $w$ corresponds to the edge of the nanoribbon parallel to the nanoscroll axis. The length $L$ for the Archimedean spiral is determined by the following equation~\cite{Siahlo2017}
\begin{align}\label{eq:01}
   L &= L(\varphi_\text{in},\varphi_\text{out}) = \int_{\varphi_\text{in}}^{\varphi_\text{out}} \frac{h}{2\pi} \sqrt{1 + \varphi^2}\,d\varphi = \notag\\
   &= \frac{h}{4\pi}\biggl(\varphi_\text{out}\sqrt{1 + \smash[b]{\varphi_\text{out}^2}} - \varphi_\text{in}\sqrt{1 + \smash[b]{\varphi_\text{in}^2}} +{} \notag\\
   &\hspace{12pt} + \ln\Bigl(\varphi_\text{out} + \sqrt{1 + \smash[b]{\varphi_\text{out}^2}}\Bigr) - \ln\Bigl(\varphi_\text{in} + \sqrt{1 + \smash[b]{\varphi_\text{in}^2}}\Bigr)\biggr),
\end{align}
where $\varphi_\text{in} = \pi R_\text{in}/h$ and $\varphi_\text{out} = \pi R_\text{out}/h$ are the inner and outer angles of the single-layer nanoscroll corresponding to its inner $R_\text{in}$ and outer $R_\text{out}$ radii, respectively (see the scheme of the single-layer nanoscroll in Fig.~3a). The interlayer spacing $h$ of both graphite \cite{Hanfland1989} and hexagonal boron nitride \cite{Duclaux1992} at room temperature is close to 0.335~nm. We use this value of the interlayer spacing $h$ for all the considered nanoscrolls.

The energy $E_{W1}$ of the interlayer interaction between adjacent turns of the single-layer nanoscroll is
proportional to the area of the layer overlap:
\begin{align}\label{eq:02}
   E_{W1} &= \frac{\varepsilon w}{S_a} L(\varphi_\text{in},\varphi_\text{out}-2\pi),
\end{align}
where $\varepsilon$ is the interlayer interaction energy per one atom of the nanoscroll, $L(\varphi_\text{in},\varphi_\text{out}-2\pi)$ is the length of the layer overlap, $S_a = 3\sqrt{3}a^2/4$ is the area per one atom, and $a$ is the bond length. As discussed in Section II, the scatter of the experimental data and results of calculations on the interlayer interaction energy $\varepsilon$ for the considered materials allow to estimate this energy with the accuracy of about 20\%, which determs the accuracy of our calculations. Since the difference between the bond lengths of graphite and hexagonal boron nitride is about 1.7\%, we neglect this difference and use the value $a=0.142$~nm of the bond length of graphite for all considered nanoscrolls. As the forces of the interlayer interaction drop drastically with increasing the separation between the interacting atoms, the macroscopic approach used here to estimate the interlayer interaction energy is adequate already for the overlap lengths which are several times greater than the interlayer spacing, i.e. for the overlap lengths above 1~nm.

The total elastic energy $E_\text{el}$ of the single-layer nanoscroll is determined in the similar manner to Eq.~(\ref{eq:01}), through the integration of
$(hwE_a/2\pi S_a) \sqrt{1 + \smash[b]{\varphi^2}}$ with respect to the angle $\varphi$, where $E_a=CK^2$ is the bending elastic energy per one atom, $K$ is the layer curvature, and the bending elastic constant $C$ is calculated in Section II. Analogously to Ref.~\onlinecite{Siahlo2017}, we use here an approximate expression for the layer curvature $K=1/R$. Thus the total elastic energy is given by the following expression:
\begin{align}\label{eq:03}
   &E_\text{el} = \frac{2\pi Cw}{hS_a}\int_{\varphi_\text{in}}^{\varphi_\text{out}} \frac{\sqrt{1 + \varphi^2}}{\varphi^2}\,d\varphi = \notag\\
   &= \frac{2\pi Cw}{hS_a}\biggl(\frac{\sqrt{1 + \smash[b]{\varphi_\text{in}^2}}}{\varphi_\text{in}} - \frac{\sqrt{1 + \smash[b]{\varphi_\text{out}^2}}}{\varphi_\text{out}} +{} \notag\\
   &\hspace{12pt} +\ln\Bigl(\varphi_\text{out} + \sqrt{1 + \smash[b]{\varphi_\text{out}^2}}\Bigr) - \ln\Bigl(\varphi_\text{in} + \sqrt{1 + \smash[b]{\varphi_\text{in}^2}}\Bigr)\biggr).
\end{align}
It should be noted that the relative difference between the total elastic energies calculated using Eq.~(\ref{eq:03}) and the exact expression for this energy \cite{Yamaletdinov2017} is less than 0.01\% even for the nanoscrolls with the smallest considered inner radius.

The potential energy $E$ of the single-layer nanoscroll is $E = E_\text{el} - E_{W1}$, where $E_\text{el}$ and $E_{W1}$ are given by Eqs.~(\ref{eq:03}) and (\ref{eq:02}), respectively.

For bilayer nanoscrolls, we use the analogous model where layers 1 and 2 lie on identical Archimedean spirals with double approximate distance $2h$ between adjacent spiral turns. These spirals have the same origins located at the nanoscroll axis and can be mapped onto each other through rotation by $\pi$ radians about this common origin (see the scheme of a bilayer nanoscroll in Fig.~3b). In such a model, the distance between adjacent layers 1 and 2 is approximately equal to the interlayer spacing $h$. It is convenient to take equations of these Archimedean spirals in the form $R_1 = h\varphi_1/\pi$ and $R_2 = h\varphi_2/\pi$, where $R_1$ and $R_2$ are the distances from layers 1 and 2, respectively, to the nanoscroll axis and angles $\varphi_1$ and $\varphi_2$ are measured from the origin of the spirals. Then the angles corresponding to the nearest points of adjacent layers (that is the points which are separated by the interlayer spacing $h$) are related by expressions $\varphi_1=\varphi_2-\pi$ and $\varphi_2=\varphi_1-\pi$ in the cases where layer 1 and layer 2, respectively, correspond to the internal layer in the considered pair of adjacent layers. The total interlayer interaction energy $E_{W2}$ of the bilayer nanoscroll is proportional to the total overlap area of adjacent layers 1 and 2 and takes the form
\begin{align}\label{eq:04}
   E_{W2} &= \frac{\varepsilon w}{S_a} [L(\varphi_\text{in1},\varphi_\text{out2}-\pi) + L(\varphi_\text{in2},\varphi_\text{out1}-\pi)],
\end{align}
where $\varphi_\text{in1} = \pi R_\text{in1}/h$ and $\varphi_\text{out1} = \pi R_\text{out1}/h$ are the inner and outer angles of layer 1 of the bilayer nanoscroll corresponding to its inner $R_\text{in1}$ and outer $R_\text{out1}$ radii, respectively, $\varphi_\text{in2} = \pi R_\text{in2}/h$ and $\varphi_\text{out2} = \pi R_\text{out2}/h$ are analogously the inner and outer angles of layer 2, respectively, (see Fig.~3b) and the lengths of the arcs of the spirals $L(\varphi_\text{in1},\varphi_\text{out2}-\pi)$ and $L(\varphi_\text{in2},\varphi_\text{out1}-\pi)$ are determined by Eq.~(\ref{eq:01}) taking into account the doubled distance $2h$ between adjacent spiral turns.

The potential energy $E$ of the bilayer nanoscroll is $E = E_\text{el1} + E_\text{el2} - E_{W2}$, where $E_{W2}$ is given by Eq.~(\ref{eq:04}), $E_\text{el1}$ and $E_\text{el2}$ are the elastic energies of layers 1 and 2, respectively,
determined by Eq.~(\ref{eq:03}) taking into account the doubled distance $2h$ between adjacent spiral turns. Note that from symmetry considerations $R_\text{in1} = R_\text{in2}$
and $R_\text{out1} = R_\text{out2}$ for bilayer nanoscrolls which consist of identical layers 1 and 2 (see Fig.~3b).

The dependences of the calculated potential energy $E$ of the nanoscrolls (per width $w = 1$~nm) on the inner radius $R_\text{in}$ for different lengths $L$ of the initial nanoribbons are shown by examples of SCNSs and BCNSs in Fig.~3c and 3d, respectively. The bending points of these dependences for the single-layer nanoscrolls correspond to the inner radius at which $\varphi_\text{out} - \varphi_\text{in} = 2\pi$ and, therefore, the adjacent layers begin to overlap and nonzero interlayer interaction energy appears. The bending points of the dependences for the bilayer nanoscrolls correspond to the inner radius at which $\varphi_\text{out1} - \varphi_\text{in1} = 2\pi$ and $\varphi_\text{out2} - \varphi_\text{in2} = 2\pi$. At this inner radius of a bilayer nanoscroll the total layer overlap length exceeds the initial nanoribbon length $L$ and the interlayer interaction energy begin to increase with decreasing the inner radius. The minimum $E_0$ of the dependence of the potential energy $E$ on the inner radius $R_\text{in}$ corresponds to the stable state of the nanoscroll. This minimum only exists for nanoscrolls formed from the nanoribbon of length $L$ which exceeds the minimal possible value $L_m$. The calculated values of the minimal possible length $L_m$ are shown in Table~\ref{tab:03}. 
\begin{table}
\tabcolsep=9.9pt
\centering
\caption{Calculated characteristic dimensions of single-layer carbon (C), single-layer  boron nitride (BN), bilayer carbon (C-C), boron nitride (BN-BN), and carbon/boron nitride (C-BN) nanoscrolls: minimal possible length $L_m$ of the initial nanoribbon for which the stable state of the nanoscroll exists and length $L_0$ of the initial nanoribbon for which the stable state of the nanoscroll is the ground state of the system.}\label{tab:03}
\smallskip
\begin{tabular}{cccccc}\hline
   &\multicolumn{2}{c}{single-layer}&\multicolumn{3}{c}{bilayer}\\\hline
   &C&BN&C-C&BN-BN&C-BN\\\hline
 $L_m$ (nm)&6.9&5.5&12.4&9.9&11.3\\
 $L_0$ (nm)&12.4&10.1&24.9&20.3&22.7\\
\hline
\end{tabular}
\end{table}
Since the elastic bending constant $C$ of graphene is greater than this constant for hexagonal boron nitride, the minimal possible length $L_m$ of SCNS is greater than this length of SBNNS. Analogously, for bilayer nanoscrolls, the minimal possible length $L_m$ decreases in the following order: BCNS $>$ CBNNS $>$ BBNNS. The minimal possible length $L_m$ of a bilayer nanoscroll is 1.8 times greater than this length for single-layer nanoscroll made of the same 2D material for both carbon and boron nitride nanoscrolls. In the framework of the macroscopic model used here, the stable single-layer nanoscroll formed from the nanoribbon of the minimal possible length $L_m$ has $\varphi_\text{out} - \varphi_\text{in} = 2\pi$ and zero overlap length of adjacent layers. The overlap length of more than 1~nm, where the estimation of the interlayer interaction energy using the macroscopic model becomes adequate, is achieved at the nanoribbon length $L \approx$ 8.5~nm both for SCNS and SBNNS.

\begin{figure}
   \centering
   \includegraphics{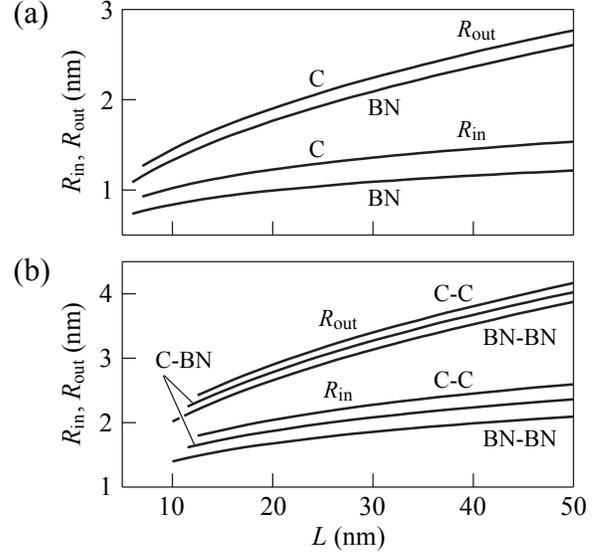}
   \caption{Calculated inner $R_\text{in}$ and outer $R_\text{out}$ radii of nanoscrolls in the stable state as functions of the length $L$ of the initial nanoribbon: (a)~single-layer carbon (C) and boron nitride (BN) nanoscrolls, (b)~bilayer carbon (C-C), boron nitride (BN-BN), and carbon/boron nitride (C-BN) nanoscrolls.}
   \label{fig:4}
\end{figure}
For bilayer carbon/boron nitride nanoscrolls, the structure of the stable state is found by numerical minimization of the potential energy as a function of inner radii $R_\text{in1}$ and $R_\text{in2}$ of layers 1 and 2, respectively. The performed calculations show that
the inner radius is smaller for boron nitride layers with the lesser value of the bending elastic constant $C$ and the outer radius is greater for graphene layers with the greater value of the bending elastic constant $C$. These smaller inner and greater outer radii are considered below as inner and outer radii of the bilayer nanoscroll made of carbon/boron nitride heterostructure.
Dependences of the inner $R_\text{in}$ and outer $R_\text{out}$ radii of the nanoscrolls in the stable state on the length $L$ of the initial nanoribbon are shown in Fig.~4 for all the considered types of nanoscrolls. Both radii $R_\text{in}$ and $R_\text{out}$ increase with increasing the length $L$. The inner radius of bilayer nanoscroll is 1.6--1.7 times greater than the inner radius of the single-layer nanoscroll made from the nanoribbons of the same length and the same 2D material. Since the elastic bending constant $C$ of graphene is greater than this constant for hexagonal boron nitride, the inner radius of SCNS is greater than this radius for the SBNNS made from the nanoribbon of the same length. Analogously, for the bilayer nanoscrolls made from the nanoribbons of the same length, the inner radius decreases in the following order: BCNS $>$ CBNNS $>$ BBNNS.

\begin{figure}
   \centering
   \includegraphics{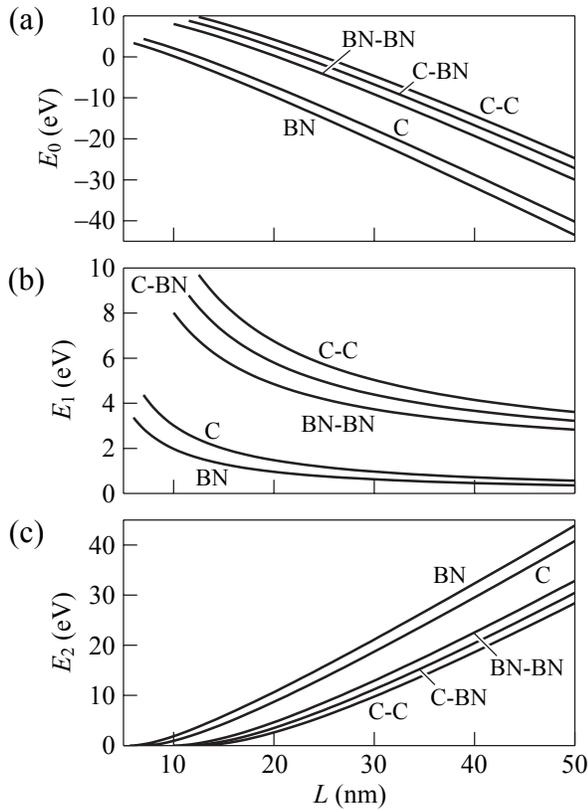}
   \caption{Calculated (a) potential energy $E_0$ in the stable state, (b)~barrier to rolling $E_1$, (c)~barrier to unrolling $E_2$ of nanoscrolls per unit width (1~nm) of the initial nanoribbon as functions of the length $L$ of the nanoribbon: single-layer carbon (C) and boron nitride (BN) and bilayer carbon (C-C), boron nitride (BN-BN), and carbon/boron nitride (C-BN) nanoscrolls. The energies of the bilayer nanoscrolls are measured relative to the energy of the corresponding flat bilayer nanoribbon.}
   \label{fig:5}
\end{figure}

Dependences of the potential energy $E_0$ which corresponds to the stable state of the nanoscroll on the length $L$ of the initial nanoribbon are shown in Fig.~5a for all the considered types of nanoscrolls. The stable state of the nanoscroll is the ground state of the system for nanoscrolls made from the nanoribbon of length $L$ which exceeds the value $L_0$. The calculated values of length $L_0$ are shown in Table~\ref{tab:03}.
The maximum of the dependence of the potential energy $E$ on the inner radius $R_\text{in}$ (see Fig.~3c and 3d) is equal to the potential barrier $E_1$ to rolling the flat nanoribbon into a nanoscroll. The barrier $E_1$ tends to zero as the length $L$ of a nanoribbon with a given width $w$ tends to infinity. The barrier to nanoscroll unrolling is $E_2 = E_1 - E_0$. This barrier tends to zero as the length $L$ of the nanoribbon with a given width $w$ tends to the minimum possible length $L_m$ for which the stable nanoscroll can exist. The dependences of barriers $E_1$ and $E_2$ on the length $L$ of the initial nanoribbon are shown in Fig.~5b and 5c, respectively, for all the considered types of the nanoscrolls.

The lifetime $\tau$ of the nanoscrolls can be estimated using the Arrhenius formula $1/\tau = \Omega\exp(-E_2/kT)$, where $\Omega$ is the frequency multiplier, $k$ is the Boltzmann constant, and $T$ is the temperature. The frequency multiplier is the same order of magnitude as the characteristic frequency of vibration corresponding to nanoscroll unrolling. We believe that such a vibration corresponds to the breathing mode of the nanoscroll which occurs with the oscillation of the overlap length \cite{Shi2009} and thus can lead to unrolling of the nanoscrolls with the smallest overlap lengths due to thermodynamic fluctuations. Let us estimate here the lifetime of such smallest nanoscrolls. Both the analytical model and molecular dynamics simulations give the frequency of breathing vibrations of the smallest SCNS made from a graphene nanoribbon of length $L = 10$~nm to be about 50 GHz \cite{Shi2009}. To estimate the order of magnitude of the nanoscroll lifetime we use the frequency multiplier $\Omega=50$ GHz for all the considered nanoscrolls. The calculated dimensions of the initial nanoribbon (length $L$ and width $w$) corresponding to nanoscrolls with the lifetime of 1000 years are shown in Fig.~6. This figure shows that if the initial nanoribbon length is only few nanometers greater than the minimal possible length $L_m$ of the stable nanoscroll then the nanoscrolls made from such initial nanoribbons are sufficiently stable for any possible applications.

\begin{figure}
   \centering
   \includegraphics{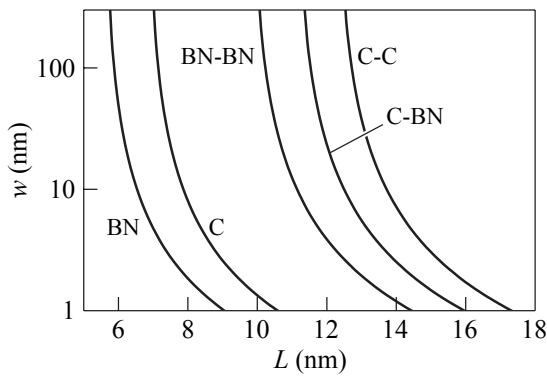}
   \caption{Calculated dimensions (length $L$ and width $w$) of the initial nanoribbons from which the nanoscrolls with the lifetime in the stable state of $\tau = 1000$ years at room temperature can be made: single-layer carbon (C) and boron nitride (BN) and bilayer carbon (C-C), boron nitride (BN-BN), and carbon/boron nitride (C-BN) nanoscrolls.}
   \label{fig:6}
\end{figure}

\section{Conclusions and discussion}

We have considered the structure and energetic characteristics of carbon, boron nitride, and carbon/boron nitride nanoscrolls made from initial single-layer and bilayer rectangular nanoribbons based on the analytical model describing the potential energy of the nanoscrolls and numerical calculations. The analytical model relies on the classical quadratic law according to which the elastic bending energy of monolayers is inversely proportional to the square of the curvature radius. While diverse data can be found in literature on the elastic energies of graphene and hexagonal boron nitride nanotubes including the cases of significant deviation from this law, the thorough analysis of the literature data and density functional theory calculations performed in the present paper convincingly demonstrate that the quadratic law matches well the dependences of the elastic energies on the curvature radius both for graphene and boron nitride layers. Therefore, the use of this law in the analytical model of nanoscrolls is completely justified.

The lengths of initial nanoribbons for which stable and energetically favorable nanoscrolls are possible correspond to the range from 7 to about 30~nm for the all considered types of nanoscrolls. The calculated barriers to rolling and unrolling of such nanoscrolls with small dimensions are within 10~eV. Simultaneously, the calculated lifetimes of nanoscrolls relative to spontaneous unrolling are greater than 1000 years at room temperature. This is so even for the nanoscrolls made from nanoribbons with the length exceeding the minimal possible length of the stable nanoscroll by only few nanometers. The calculated values of the barriers and lifetimes show that such nanoscrolls offer promise as movable parts of nanoelectromechanical systems based on rolling and unrolling of the nanoscrolls under the action of the electrostatic force analogously to already realized systems based on relative motion of walls or bending of carbon nanotubes (see Ref.~\onlinecite{Bichoutskaia2008} for a review).

\begin{acknowledgements}
AL, AP, and YL acknowledge the Russian Foundation of Basic Research (16-52-00181). IL acknowledges the financial support from Grupos Consolidados UPV/EHU del Gobierno Vasco (IT578-13) and EU-H2020 project ``MOSTOPHOS" (n. 646259). AS, NP, and SV acknowledge the Belarusian Republican Foundation for Fundamental Research (grant No.~F16R-107) and Belarusian National Research Program ``Convergence''. This work has been carried out using computing resources of the federal collective usage center Complex for Simulation and Data Processing for Mega-science Facilities at NRC ``Kurchatov Institute'', http://ckp.nrcki.ru/.
\end{acknowledgements}

\bibliography{rsc}

\begin{thebibliography}{84}%
\makeatletter
\providecommand \@ifxundefined [1]{%
 \@ifx{#1\undefined}
}%
\providecommand \@ifnum [1]{%
 \ifnum #1\expandafter \@firstoftwo
 \else \expandafter \@secondoftwo
 \fi
}%
\providecommand \@ifx [1]{%
 \ifx #1\expandafter \@firstoftwo
 \else \expandafter \@secondoftwo
 \fi
}%
\providecommand \natexlab [1]{#1}%
\providecommand \enquote  [1]{``#1''}%
\providecommand \bibnamefont  [1]{#1}%
\providecommand \bibfnamefont [1]{#1}%
\providecommand \citenamefont [1]{#1}%
\providecommand \href@noop [0]{\@secondoftwo}%
\providecommand \href [0]{\begingroup \@sanitize@url \@href}%
\providecommand \@href[1]{\@@startlink{#1}\@@href}%
\providecommand \@@href[1]{\endgroup#1\@@endlink}%
\providecommand \@sanitize@url [0]{\catcode `\\12\catcode `\$12\catcode
  `\&12\catcode `\#12\catcode `\^12\catcode `\_12\catcode `\%12\relax}%
\providecommand \@@startlink[1]{}%
\providecommand \@@endlink[0]{}%
\providecommand \url  [0]{\begingroup\@sanitize@url \@url }%
\providecommand \@url [1]{\endgroup\@href {#1}{\urlprefix }}%
\providecommand \urlprefix  [0]{URL }%
\providecommand \Eprint [0]{\href }%
\providecommand \doibase [0]{http://dx.doi.org/}%
\providecommand \selectlanguage [0]{\@gobble}%
\providecommand \bibinfo  [0]{\@secondoftwo}%
\providecommand \bibfield  [0]{\@secondoftwo}%
\providecommand \translation [1]{[#1]}%
\providecommand \BibitemOpen [0]{}%
\providecommand \bibitemStop [0]{}%
\providecommand \bibitemNoStop [0]{.\EOS\space}%
\providecommand \EOS [0]{\spacefactor3000\relax}%
\providecommand \BibitemShut  [1]{\csname bibitem#1\endcsname}%
\let\auto@bib@innerbib\@empty
\bibitem [{\citenamefont {Viculis}\ \emph {et~al.}(2003)\citenamefont
  {Viculis}, \citenamefont {Mack},\ and\ \citenamefont {Kaner}}]{Viculis2003}%
  \BibitemOpen
  \bibfield  {author} {\bibinfo {author} {\bibfnamefont {L.~M.}\ \bibnamefont
  {Viculis}}, \bibinfo {author} {\bibfnamefont {J.~J.}\ \bibnamefont {Mack}}, \
  and\ \bibinfo {author} {\bibfnamefont {R.~B.}\ \bibnamefont {Kaner}},\ }\href
  {\doibase 10.1126/science.1078842} {\bibfield  {journal} {\bibinfo  {journal}
  {Science}\ }\textbf {\bibinfo {volume} {299}},\ \bibinfo {pages} {1361}
  (\bibinfo {year} {2003})}\BibitemShut {NoStop}%
\bibitem [{\citenamefont {Shioyama}\ and\ \citenamefont
  {Akita}(2003)}]{Shioyama2003}%
  \BibitemOpen
  \bibfield  {author} {\bibinfo {author} {\bibfnamefont {H.}~\bibnamefont
  {Shioyama}}\ and\ \bibinfo {author} {\bibfnamefont {T.}~\bibnamefont
  {Akita}},\ }\href {\doibase 10.1016/S0008-6223(02)00278-6} {\bibfield
  {journal} {\bibinfo  {journal} {Carbon}\ }\textbf {\bibinfo {volume} {41}},\
  \bibinfo {pages} {179} (\bibinfo {year} {2003})}\BibitemShut {NoStop}%
\bibitem [{\citenamefont {Savoskin}\ \emph {et~al.}(2007)\citenamefont
  {Savoskin}, \citenamefont {Mochalin}, \citenamefont {Yaroshenko},
  \citenamefont {Lazareva}, \citenamefont {Konstantinova}, \citenamefont
  {Barsukov},\ and\ \citenamefont {Prokofiev}}]{Savoskin2007}%
  \BibitemOpen
  \bibfield  {author} {\bibinfo {author} {\bibfnamefont {M.~V.}\ \bibnamefont
  {Savoskin}}, \bibinfo {author} {\bibfnamefont {V.~N.}\ \bibnamefont
  {Mochalin}}, \bibinfo {author} {\bibfnamefont {A.~P.}\ \bibnamefont
  {Yaroshenko}}, \bibinfo {author} {\bibfnamefont {N.~I.}\ \bibnamefont
  {Lazareva}}, \bibinfo {author} {\bibfnamefont {T.~E.}\ \bibnamefont
  {Konstantinova}}, \bibinfo {author} {\bibfnamefont {I.~V.}\ \bibnamefont
  {Barsukov}}, \ and\ \bibinfo {author} {\bibfnamefont {I.~G.}\ \bibnamefont
  {Prokofiev}},\ }\href {\doibase 10.1016/j.carbon.2007.09.031} {\bibfield
  {journal} {\bibinfo  {journal} {Carbon}\ }\textbf {\bibinfo {volume} {45}},\
  \bibinfo {pages} {2797} (\bibinfo {year} {2007})}\BibitemShut {NoStop}%
\bibitem [{\citenamefont {Li}\ \emph {et~al.}(2005)\citenamefont {Li},
  \citenamefont {Peng}, \citenamefont {Bai},\ and\ \citenamefont
  {Jiang}}]{Li2005}%
  \BibitemOpen
  \bibfield  {author} {\bibinfo {author} {\bibfnamefont {J.~L.}\ \bibnamefont
  {Li}}, \bibinfo {author} {\bibfnamefont {Q.~S.}\ \bibnamefont {Peng}},
  \bibinfo {author} {\bibfnamefont {G.~Z.}\ \bibnamefont {Bai}}, \ and\
  \bibinfo {author} {\bibfnamefont {W.}~\bibnamefont {Jiang}},\ }\href
  {\doibase 10.1016/j.carbon.2005.06.007} {\bibfield  {journal} {\bibinfo
  {journal} {Carbon}\ }\textbf {\bibinfo {volume} {43}},\ \bibinfo {pages}
  {2830} (\bibinfo {year} {2005})}\BibitemShut {NoStop}%
\bibitem [{\citenamefont {Ruland}\ \emph {et~al.}(2003)\citenamefont {Ruland},
  \citenamefont {Schaper}, \citenamefont {Hou},\ and\ \citenamefont
  {Greiner}}]{Ruland2003}%
  \BibitemOpen
  \bibfield  {author} {\bibinfo {author} {\bibfnamefont {W.}~\bibnamefont
  {Ruland}}, \bibinfo {author} {\bibfnamefont {A.~K.}\ \bibnamefont {Schaper}},
  \bibinfo {author} {\bibfnamefont {H.}~\bibnamefont {Hou}}, \ and\ \bibinfo
  {author} {\bibfnamefont {A.}~\bibnamefont {Greiner}},\ }\href {\doibase
  10.1016/S0008-6223(02)00342-1} {\bibfield  {journal} {\bibinfo  {journal}
  {Carbon}\ }\textbf {\bibinfo {volume} {41}},\ \bibinfo {pages} {423}
  (\bibinfo {year} {2003})}\BibitemShut {NoStop}%
\bibitem [{\citenamefont {Chuvilin}\ \emph {et~al.}(2009)\citenamefont
  {Chuvilin}, \citenamefont {Kuznetsov},\ and\ \citenamefont
  {Obraztsov}}]{Chuvilin2009}%
  \BibitemOpen
  \bibfield  {author} {\bibinfo {author} {\bibfnamefont {A.~L.}\ \bibnamefont
  {Chuvilin}}, \bibinfo {author} {\bibfnamefont {V.~L.}\ \bibnamefont
  {Kuznetsov}}, \ and\ \bibinfo {author} {\bibfnamefont {A.~N.}\ \bibnamefont
  {Obraztsov}},\ }\href {\doibase 10.1016/j.carbon.2009.07.024} {\bibfield
  {journal} {\bibinfo  {journal} {Carbon}\ }\textbf {\bibinfo {volume} {47}},\
  \bibinfo {pages} {3099} (\bibinfo {year} {2009})}\BibitemShut {NoStop}%
\bibitem [{\citenamefont {Schaper}\ \emph {et~al.}(2011)\citenamefont
  {Schaper}, \citenamefont {Wang}, \citenamefont {Xu}, \citenamefont {Bando},\
  and\ \citenamefont {Golberg}}]{Schaper2011}%
  \BibitemOpen
  \bibfield  {author} {\bibinfo {author} {\bibfnamefont {A.~K.}\ \bibnamefont
  {Schaper}}, \bibinfo {author} {\bibfnamefont {M.~S.}\ \bibnamefont {Wang}},
  \bibinfo {author} {\bibfnamefont {Z.}~\bibnamefont {Xu}}, \bibinfo {author}
  {\bibfnamefont {Y.}~\bibnamefont {Bando}}, \ and\ \bibinfo {author}
  {\bibfnamefont {D.}~\bibnamefont {Golberg}},\ }\href {\doibase
  10.1021/nl201655c} {\bibfield  {journal} {\bibinfo  {journal} {Nano Lett.}\
  }\textbf {\bibinfo {volume} {11}},\ \bibinfo {pages} {3295} (\bibinfo {year}
  {2011})}\BibitemShut {NoStop}%
\bibitem [{\citenamefont {Sidorov}\ \emph {et~al.}(2009)\citenamefont
  {Sidorov}, \citenamefont {Mudd}, \citenamefont {Sumanasekera}, \citenamefont
  {Ouseph}, \citenamefont {Jayanthi},\ and\ \citenamefont {Wu}}]{Sidorov2009}%
  \BibitemOpen
  \bibfield  {author} {\bibinfo {author} {\bibfnamefont {A.}~\bibnamefont
  {Sidorov}}, \bibinfo {author} {\bibfnamefont {D.}~\bibnamefont {Mudd}},
  \bibinfo {author} {\bibfnamefont {G.}~\bibnamefont {Sumanasekera}}, \bibinfo
  {author} {\bibfnamefont {P.~J.}\ \bibnamefont {Ouseph}}, \bibinfo {author}
  {\bibfnamefont {C.~S.}\ \bibnamefont {Jayanthi}}, \ and\ \bibinfo {author}
  {\bibfnamefont {S.-Y.}\ \bibnamefont {Wu}},\ }\href {\doibase
  10.1088/0957-4484/20/5/055611} {\bibfield  {journal} {\bibinfo  {journal}
  {Nanotechnology}\ }\textbf {\bibinfo {volume} {20}},\ \bibinfo {pages}
  {055611} (\bibinfo {year} {2009})}\BibitemShut {NoStop}%
\bibitem [{\citenamefont {Zeng}\ \emph {et~al.}(2011)\citenamefont {Zeng},
  \citenamefont {Kuang}, \citenamefont {Wang}, \citenamefont {Huang},
  \citenamefont {Fu},\ and\ \citenamefont {Zhou}}]{Zeng2011}%
  \BibitemOpen
  \bibfield  {author} {\bibinfo {author} {\bibfnamefont {F.}~\bibnamefont
  {Zeng}}, \bibinfo {author} {\bibfnamefont {Y.}~\bibnamefont {Kuang}},
  \bibinfo {author} {\bibfnamefont {Y.}~\bibnamefont {Wang}}, \bibinfo {author}
  {\bibfnamefont {Z.}~\bibnamefont {Huang}}, \bibinfo {author} {\bibfnamefont
  {C.}~\bibnamefont {Fu}}, \ and\ \bibinfo {author} {\bibfnamefont
  {H.}~\bibnamefont {Zhou}},\ }\href {\doibase 10.1002/adma.201102798}
  {\bibfield  {journal} {\bibinfo  {journal} {Adv. Mater.}\ }\textbf {\bibinfo
  {volume} {23}},\ \bibinfo {pages} {4929} (\bibinfo {year}
  {2011})}\BibitemShut {NoStop}%
\bibitem [{\citenamefont {Zeng}\ \emph {et~al.}(2012)\citenamefont {Zeng},
  \citenamefont {Kuang}, \citenamefont {Liu}, \citenamefont {Liu},
  \citenamefont {Huang}, \citenamefont {Fu},\ and\ \citenamefont
  {Zhou}}]{Zeng2012}%
  \BibitemOpen
  \bibfield  {author} {\bibinfo {author} {\bibfnamefont {F.}~\bibnamefont
  {Zeng}}, \bibinfo {author} {\bibfnamefont {Y.}~\bibnamefont {Kuang}},
  \bibinfo {author} {\bibfnamefont {G.}~\bibnamefont {Liu}}, \bibinfo {author}
  {\bibfnamefont {R.}~\bibnamefont {Liu}}, \bibinfo {author} {\bibfnamefont
  {Z.}~\bibnamefont {Huang}}, \bibinfo {author} {\bibfnamefont
  {C.}~\bibnamefont {Fu}}, \ and\ \bibinfo {author} {\bibfnamefont
  {H.}~\bibnamefont {Zhou}},\ }\href {\doibase 10.1039/C2NR30779K} {\bibfield
  {journal} {\bibinfo  {journal} {Nanoscale}\ }\textbf {\bibinfo {volume}
  {4}},\ \bibinfo {pages} {3997} (\bibinfo {year} {2012})}\BibitemShut
  {NoStop}%
\bibitem [{\citenamefont {Zheng}\ \emph {et~al.}(2011)\citenamefont {Zheng},
  \citenamefont {Liu}, \citenamefont {Wu}, \citenamefont {Guo}, \citenamefont
  {Wu}, \citenamefont {Yu}, \citenamefont {Liu},\ and\ \citenamefont
  {Zhu}}]{Zheng2011}%
  \BibitemOpen
  \bibfield  {author} {\bibinfo {author} {\bibfnamefont {J.}~\bibnamefont
  {Zheng}}, \bibinfo {author} {\bibfnamefont {H.}~\bibnamefont {Liu}}, \bibinfo
  {author} {\bibfnamefont {B.}~\bibnamefont {Wu}}, \bibinfo {author}
  {\bibfnamefont {Y.}~\bibnamefont {Guo}}, \bibinfo {author} {\bibfnamefont
  {T.}~\bibnamefont {Wu}}, \bibinfo {author} {\bibfnamefont {G.}~\bibnamefont
  {Yu}}, \bibinfo {author} {\bibfnamefont {Y.}~\bibnamefont {Liu}}, \ and\
  \bibinfo {author} {\bibfnamefont {D.}~\bibnamefont {Zhu}},\ }\href {\doibase
  10.1002/adma.201004759} {\bibfield  {journal} {\bibinfo  {journal} {Adv.
  Mater.}\ }\textbf {\bibinfo {volume} {23}},\ \bibinfo {pages} {2460}
  (\bibinfo {year} {2011})}\BibitemShut {NoStop}%
\bibitem [{\citenamefont {Xie}\ \emph {et~al.}(2009)\citenamefont {Xie},
  \citenamefont {Ju}, \citenamefont {Feng}, \citenamefont {Sun}, \citenamefont
  {Zhou}, \citenamefont {Liu}, \citenamefont {Fan}, \citenamefont {Li},\ and\
  \citenamefont {Jiang}}]{Xie2009}%
  \BibitemOpen
  \bibfield  {author} {\bibinfo {author} {\bibfnamefont {X.}~\bibnamefont
  {Xie}}, \bibinfo {author} {\bibfnamefont {L.}~\bibnamefont {Ju}}, \bibinfo
  {author} {\bibfnamefont {X.}~\bibnamefont {Feng}}, \bibinfo {author}
  {\bibfnamefont {Y.}~\bibnamefont {Sun}}, \bibinfo {author} {\bibfnamefont
  {R.}~\bibnamefont {Zhou}}, \bibinfo {author} {\bibfnamefont {K.}~\bibnamefont
  {Liu}}, \bibinfo {author} {\bibfnamefont {S.}~\bibnamefont {Fan}}, \bibinfo
  {author} {\bibfnamefont {Q.}~\bibnamefont {Li}}, \ and\ \bibinfo {author}
  {\bibfnamefont {K.}~\bibnamefont {Jiang}},\ }\href {\doibase
  10.1021/nl900677y} {\bibfield  {journal} {\bibinfo  {journal} {Nano Lett.}\
  }\textbf {\bibinfo {volume} {9}},\ \bibinfo {pages} {2565} (\bibinfo {year}
  {2009})}\BibitemShut {NoStop}%
\bibitem [{\citenamefont {Mirsaidov}\ \emph {et~al.}(2013)\citenamefont
  {Mirsaidov}, \citenamefont {Mokkapati}, \citenamefont {Bhattacharya},
  \citenamefont {Andersen}, \citenamefont {Bosman}, \citenamefont
  {{\"O}zyilmaz},\ and\ \citenamefont {Matsudaira}}]{Mirsaidov2013}%
  \BibitemOpen
  \bibfield  {author} {\bibinfo {author} {\bibfnamefont {U.}~\bibnamefont
  {Mirsaidov}}, \bibinfo {author} {\bibfnamefont {V.~R. S.~S.}\ \bibnamefont
  {Mokkapati}}, \bibinfo {author} {\bibfnamefont {D.}~\bibnamefont
  {Bhattacharya}}, \bibinfo {author} {\bibfnamefont {H.}~\bibnamefont
  {Andersen}}, \bibinfo {author} {\bibfnamefont {M.}~\bibnamefont {Bosman}},
  \bibinfo {author} {\bibfnamefont {B.}~\bibnamefont {{\"O}zyilmaz}}, \ and\
  \bibinfo {author} {\bibfnamefont {P.}~\bibnamefont {Matsudaira}},\ }\href
  {\doibase 10.1039/C3LC50304F} {\bibfield  {journal} {\bibinfo  {journal} {Lab
  Chip}\ }\textbf {\bibinfo {volume} {13}},\ \bibinfo {pages} {2874} (\bibinfo
  {year} {2013})}\BibitemShut {NoStop}%
\bibitem [{\citenamefont {Yan}\ \emph {et~al.}(2013)\citenamefont {Yan},
  \citenamefont {Wang}, \citenamefont {Han}, \citenamefont {Ma}, \citenamefont
  {Xu}, \citenamefont {An}, \citenamefont {Xu}, \citenamefont {Niu},
  \citenamefont {Zhao}, \citenamefont {Tian}, \citenamefont {Hu}, \citenamefont
  {Wu},\ and\ \citenamefont {Mai}}]{Yan2013}%
  \BibitemOpen
  \bibfield  {author} {\bibinfo {author} {\bibfnamefont {M.}~\bibnamefont
  {Yan}}, \bibinfo {author} {\bibfnamefont {F.}~\bibnamefont {Wang}}, \bibinfo
  {author} {\bibfnamefont {C.}~\bibnamefont {Han}}, \bibinfo {author}
  {\bibfnamefont {X.}~\bibnamefont {Ma}}, \bibinfo {author} {\bibfnamefont
  {X.}~\bibnamefont {Xu}}, \bibinfo {author} {\bibfnamefont {Q.}~\bibnamefont
  {An}}, \bibinfo {author} {\bibfnamefont {L.}~\bibnamefont {Xu}}, \bibinfo
  {author} {\bibfnamefont {C.}~\bibnamefont {Niu}}, \bibinfo {author}
  {\bibfnamefont {Y.}~\bibnamefont {Zhao}}, \bibinfo {author} {\bibfnamefont
  {X.}~\bibnamefont {Tian}}, \bibinfo {author} {\bibfnamefont {P.}~\bibnamefont
  {Hu}}, \bibinfo {author} {\bibfnamefont {H.}~\bibnamefont {Wu}}, \ and\
  \bibinfo {author} {\bibfnamefont {L.}~\bibnamefont {Mai}},\ }\href {\doibase
  10.1021/ja409027s} {\bibfield  {journal} {\bibinfo  {journal} {J. Am. Chem.
  Soc.}\ }\textbf {\bibinfo {volume} {135}},\ \bibinfo {pages} {18176}
  (\bibinfo {year} {2013})}\BibitemShut {NoStop}%
\bibitem [{\citenamefont {Zheng}\ \emph {et~al.}(2016)\citenamefont {Zheng},
  \citenamefont {Xu},\ and\ \citenamefont {Gao}}]{Zheng2016}%
  \BibitemOpen
  \bibfield  {author} {\bibinfo {author} {\bibfnamefont {B.}~\bibnamefont
  {Zheng}}, \bibinfo {author} {\bibfnamefont {Z.}~\bibnamefont {Xu}}, \ and\
  \bibinfo {author} {\bibfnamefont {C.}~\bibnamefont {Gao}},\ }\href {\doibase
  10.1039/C5NR07067H} {\bibfield  {journal} {\bibinfo  {journal} {Nanoscale}\
  }\textbf {\bibinfo {volume} {8}},\ \bibinfo {pages} {1413} (\bibinfo {year}
  {2016})}\BibitemShut {NoStop}%
\bibitem [{\citenamefont {Pan}\ \emph {et~al.}(2005)\citenamefont {Pan},
  \citenamefont {Feng},\ and\ \citenamefont {Lin}}]{Pan2005}%
  \BibitemOpen
  \bibfield  {author} {\bibinfo {author} {\bibfnamefont {H.}~\bibnamefont
  {Pan}}, \bibinfo {author} {\bibfnamefont {Y.}~\bibnamefont {Feng}}, \ and\
  \bibinfo {author} {\bibfnamefont {J.}~\bibnamefont {Lin}},\ }\href {\doibase
  10.1103/PhysRevB.72.085415} {\bibfield  {journal} {\bibinfo  {journal} {Phys.
  Rev. B}\ }\textbf {\bibinfo {volume} {72}},\ \bibinfo {pages} {085415}
  (\bibinfo {year} {2005})}\BibitemShut {NoStop}%
\bibitem [{\citenamefont {Rurali}\ \emph {et~al.}(2006)\citenamefont {Rurali},
  \citenamefont {Coluci},\ and\ \citenamefont {Galv\~ao}}]{Rurali2006}%
  \BibitemOpen
  \bibfield  {author} {\bibinfo {author} {\bibfnamefont {R.}~\bibnamefont
  {Rurali}}, \bibinfo {author} {\bibfnamefont {V.~R.}\ \bibnamefont {Coluci}},
  \ and\ \bibinfo {author} {\bibfnamefont {D.~S.}\ \bibnamefont {Galv\~ao}},\
  }\href {\doibase 10.1103/PhysRevB.74.085414} {\bibfield  {journal} {\bibinfo
  {journal} {Phys. Rev. B}\ }\textbf {\bibinfo {volume} {74}},\ \bibinfo
  {pages} {085414} (\bibinfo {year} {2006})}\BibitemShut {NoStop}%
\bibitem [{\citenamefont {Chen}\ \emph {et~al.}(2007)\citenamefont {Chen},
  \citenamefont {Lu},\ and\ \citenamefont {Gao}}]{Chen2007}%
  \BibitemOpen
  \bibfield  {author} {\bibinfo {author} {\bibfnamefont {Y.}~\bibnamefont
  {Chen}}, \bibinfo {author} {\bibfnamefont {J.}~\bibnamefont {Lu}}, \ and\
  \bibinfo {author} {\bibfnamefont {Z.}~\bibnamefont {Gao}},\ }\href {\doibase
  10.1021/jp066030r} {\bibfield  {journal} {\bibinfo  {journal} {J. Phys. Chem.
  C}\ }\textbf {\bibinfo {volume} {111}},\ \bibinfo {pages} {1625} (\bibinfo
  {year} {2007})}\BibitemShut {NoStop}%
\bibitem [{\citenamefont {Dong}\ \emph {et~al.}(2016)\citenamefont {Dong},
  \citenamefont {Zhang}, \citenamefont {Fang}, \citenamefont {Gong},
  \citenamefont {Zhang},\ and\ \citenamefont {Zhang}}]{Dong2016}%
  \BibitemOpen
  \bibfield  {author} {\bibinfo {author} {\bibfnamefont {H.}~\bibnamefont
  {Dong}}, \bibinfo {author} {\bibfnamefont {Y.}~\bibnamefont {Zhang}},
  \bibinfo {author} {\bibfnamefont {D.}~\bibnamefont {Fang}}, \bibinfo {author}
  {\bibfnamefont {B.}~\bibnamefont {Gong}}, \bibinfo {author} {\bibfnamefont
  {E.}~\bibnamefont {Zhang}}, \ and\ \bibinfo {author} {\bibfnamefont
  {S.}~\bibnamefont {Zhang}},\ }\href {\doibase 10.1039/C5NR07628E} {\bibfield
  {journal} {\bibinfo  {journal} {Nanoscale}\ }\textbf {\bibinfo {volume}
  {8}},\ \bibinfo {pages} {2887} (\bibinfo {year} {2016})}\BibitemShut
  {NoStop}%
\bibitem [{\citenamefont {Karimi}\ \emph {et~al.}(2014)\citenamefont {Karimi},
  \citenamefont {Ahmadi}, \citenamefont {Khosrowabadi}, \citenamefont
  {Rahmani}, \citenamefont {Saeidimanesh}, \citenamefont {Ismail},
  \citenamefont {Naghib},\ and\ \citenamefont {Akbari}}]{Karimi2014}%
  \BibitemOpen
  \bibfield  {author} {\bibinfo {author} {\bibfnamefont {H.}~\bibnamefont
  {Karimi}}, \bibinfo {author} {\bibfnamefont {M.~T.}\ \bibnamefont {Ahmadi}},
  \bibinfo {author} {\bibfnamefont {E.}~\bibnamefont {Khosrowabadi}}, \bibinfo
  {author} {\bibfnamefont {R.}~\bibnamefont {Rahmani}}, \bibinfo {author}
  {\bibfnamefont {M.}~\bibnamefont {Saeidimanesh}}, \bibinfo {author}
  {\bibfnamefont {R.}~\bibnamefont {Ismail}}, \bibinfo {author} {\bibfnamefont
  {S.~D.}\ \bibnamefont {Naghib}}, \ and\ \bibinfo {author} {\bibfnamefont
  {E.}~\bibnamefont {Akbari}},\ }\href {\doibase 10.1039/C3RA47432A} {\bibfield
   {journal} {\bibinfo  {journal} {RSC Adv.}\ }\textbf {\bibinfo {volume}
  {4}},\ \bibinfo {pages} {16153} (\bibinfo {year} {2014})}\BibitemShut
  {NoStop}%
\bibitem [{\citenamefont {Khaledian}\ \emph {et~al.}(2015)\citenamefont
  {Khaledian}, \citenamefont {Ismail}, \citenamefont {Saeidmanesh},\ and\
  \citenamefont {Khaledian}}]{Khaledian2015}%
  \BibitemOpen
  \bibfield  {author} {\bibinfo {author} {\bibfnamefont {M.}~\bibnamefont
  {Khaledian}}, \bibinfo {author} {\bibfnamefont {R.}~\bibnamefont {Ismail}},
  \bibinfo {author} {\bibfnamefont {M.}~\bibnamefont {Saeidmanesh}}, \ and\
  \bibinfo {author} {\bibfnamefont {P.}~\bibnamefont {Khaledian}},\ }\href
  {\doibase 10.1039/C5RA01150G} {\bibfield  {journal} {\bibinfo  {journal} {RSC
  Adv.}\ }\textbf {\bibinfo {volume} {5}},\ \bibinfo {pages} {54700} (\bibinfo
  {year} {2015})}\BibitemShut {NoStop}%
\bibitem [{\citenamefont {Shi}\ \emph {et~al.}(2010{\natexlab{a}})\citenamefont
  {Shi}, \citenamefont {Cheng}, \citenamefont {Pugno},\ and\ \citenamefont
  {Gao}}]{Shi2010}%
  \BibitemOpen
  \bibfield  {author} {\bibinfo {author} {\bibfnamefont {X.}~\bibnamefont
  {Shi}}, \bibinfo {author} {\bibfnamefont {Y.}~\bibnamefont {Cheng}}, \bibinfo
  {author} {\bibfnamefont {N.~M.}\ \bibnamefont {Pugno}}, \ and\ \bibinfo
  {author} {\bibfnamefont {H.}~\bibnamefont {Gao}},\ }\href {\doibase
  10.1002/smll.200902286} {\bibfield  {journal} {\bibinfo  {journal} {Small}\
  }\textbf {\bibinfo {volume} {6}},\ \bibinfo {pages} {739} (\bibinfo {year}
  {2010}{\natexlab{a}})}\BibitemShut {NoStop}%
\bibitem [{\citenamefont {Mpourmpakis}\ \emph {et~al.}(2007)\citenamefont
  {Mpourmpakis}, \citenamefont {Tylianakis},\ and\ \citenamefont
  {Froudakis}}]{Mpourmpakis2007}%
  \BibitemOpen
  \bibfield  {author} {\bibinfo {author} {\bibfnamefont {G.}~\bibnamefont
  {Mpourmpakis}}, \bibinfo {author} {\bibfnamefont {E.}~\bibnamefont
  {Tylianakis}}, \ and\ \bibinfo {author} {\bibfnamefont {G.~E.}\ \bibnamefont
  {Froudakis}},\ }\href {\doibase 10.1021/nl070530u} {\bibfield  {journal}
  {\bibinfo  {journal} {Nano Lett.}\ }\textbf {\bibinfo {volume} {7}},\
  \bibinfo {pages} {1893} (\bibinfo {year} {2007})}\BibitemShut {NoStop}%
\bibitem [{\citenamefont {Coluci}\ \emph {et~al.}(2007)\citenamefont {Coluci},
  \citenamefont {Braga}, \citenamefont {Baughman},\ and\ \citenamefont
  {Galv\~ao}}]{Coluci2007}%
  \BibitemOpen
  \bibfield  {author} {\bibinfo {author} {\bibfnamefont {V.~R.}\ \bibnamefont
  {Coluci}}, \bibinfo {author} {\bibfnamefont {S.~F.}\ \bibnamefont {Braga}},
  \bibinfo {author} {\bibfnamefont {R.~H.}\ \bibnamefont {Baughman}}, \ and\
  \bibinfo {author} {\bibfnamefont {D.~S.}\ \bibnamefont {Galv\~ao}},\ }\href
  {\doibase 10.1103/PhysRevB.75.125404} {\bibfield  {journal} {\bibinfo
  {journal} {Phys. Rev. B}\ }\textbf {\bibinfo {volume} {75}},\ \bibinfo
  {pages} {125404} (\bibinfo {year} {2007})}\BibitemShut {NoStop}%
\bibitem [{\citenamefont {Braga}\ \emph {et~al.}(2007)\citenamefont {Braga},
  \citenamefont {Coluci}, \citenamefont {Baughman},\ and\ \citenamefont
  {Galv\~ao}}]{Braga2007}%
  \BibitemOpen
  \bibfield  {author} {\bibinfo {author} {\bibfnamefont {S.~F.}\ \bibnamefont
  {Braga}}, \bibinfo {author} {\bibfnamefont {V.~R.}\ \bibnamefont {Coluci}},
  \bibinfo {author} {\bibfnamefont {R.~H.}\ \bibnamefont {Baughman}}, \ and\
  \bibinfo {author} {\bibfnamefont {D.~S.}\ \bibnamefont {Galv\~ao}},\ }\href
  {\doibase https://doi.org/10.1016/j.cplett.2007.04.060} {\bibfield  {journal}
  {\bibinfo  {journal} {Chem. Phys. Lett.}\ }\textbf {\bibinfo {volume}
  {441}},\ \bibinfo {pages} {78} (\bibinfo {year} {2007})}\BibitemShut
  {NoStop}%
\bibitem [{\citenamefont {Ceperley}\ and\ \citenamefont
  {Alder}(1980)}]{Ceperley1980}%
  \BibitemOpen
  \bibfield  {author} {\bibinfo {author} {\bibfnamefont {D.~M.}\ \bibnamefont
  {Ceperley}}\ and\ \bibinfo {author} {\bibfnamefont {B.~J.}\ \bibnamefont
  {Alder}},\ }\href {\doibase 10.1103/PhysRevLett.45.566} {\bibfield  {journal}
  {\bibinfo  {journal} {Phys. Rev. Lett.}\ }\textbf {\bibinfo {volume} {45}},\
  \bibinfo {pages} {566} (\bibinfo {year} {1980})}\BibitemShut {NoStop}%
\bibitem [{\citenamefont {Lebedeva}\ \emph {et~al.}(2010)\citenamefont
  {Lebedeva}, \citenamefont {Knizhnik}, \citenamefont {Popov}, \citenamefont
  {Ershova}, \citenamefont {Lozovik},\ and\ \citenamefont
  {Potapkin}}]{Lebedeva2010}%
  \BibitemOpen
  \bibfield  {author} {\bibinfo {author} {\bibfnamefont {I.~V.}\ \bibnamefont
  {Lebedeva}}, \bibinfo {author} {\bibfnamefont {A.~A.}\ \bibnamefont
  {Knizhnik}}, \bibinfo {author} {\bibfnamefont {A.~M.}\ \bibnamefont {Popov}},
  \bibinfo {author} {\bibfnamefont {O.~V.}\ \bibnamefont {Ershova}}, \bibinfo
  {author} {\bibfnamefont {Y.~E.}\ \bibnamefont {Lozovik}}, \ and\ \bibinfo
  {author} {\bibfnamefont {B.~V.}\ \bibnamefont {Potapkin}},\ }\href {\doibase
  10.1103/PhysRevB.82.155460} {\bibfield  {journal} {\bibinfo  {journal} {Phys.
  Rev. B}\ }\textbf {\bibinfo {volume} {82}},\ \bibinfo {pages} {155460}
  (\bibinfo {year} {2010})}\BibitemShut {NoStop}%
\bibitem [{\citenamefont {Kolmogorov}\ and\ \citenamefont
  {Crespi}(2005)}]{Kolmogorov2005}%
  \BibitemOpen
  \bibfield  {author} {\bibinfo {author} {\bibfnamefont {A.~N.}\ \bibnamefont
  {Kolmogorov}}\ and\ \bibinfo {author} {\bibfnamefont {V.~H.}\ \bibnamefont
  {Crespi}},\ }\href {\doibase 10.1103/PhysRevB.71.235415} {\bibfield
  {journal} {\bibinfo  {journal} {Phys. Rev. B}\ }\textbf {\bibinfo {volume}
  {71}},\ \bibinfo {pages} {235415} (\bibinfo {year} {2005})}\BibitemShut
  {NoStop}%
\bibitem [{\citenamefont {Wang}\ \emph {et~al.}(2014)\citenamefont {Wang},
  \citenamefont {Selbach},\ and\ \citenamefont {Grande}}]{Wang2014}%
  \BibitemOpen
  \bibfield  {author} {\bibinfo {author} {\bibfnamefont {Z.}~\bibnamefont
  {Wang}}, \bibinfo {author} {\bibfnamefont {S.~M.}\ \bibnamefont {Selbach}}, \
  and\ \bibinfo {author} {\bibfnamefont {T.}~\bibnamefont {Grande}},\ }\href
  {\doibase 10.1039/C3RA47187J} {\bibfield  {journal} {\bibinfo  {journal} {RSC
  Adv.}\ }\textbf {\bibinfo {volume} {4}},\ \bibinfo {pages} {4069} (\bibinfo
  {year} {2014})}\BibitemShut {NoStop}%
\bibitem [{\citenamefont {Reguzzoni}\ \emph {et~al.}(2012)\citenamefont
  {Reguzzoni}, \citenamefont {Fasolino}, \citenamefont {Molinari},\ and\
  \citenamefont {Righi}}]{Reguzzoni2012}%
  \BibitemOpen
  \bibfield  {author} {\bibinfo {author} {\bibfnamefont {M.}~\bibnamefont
  {Reguzzoni}}, \bibinfo {author} {\bibfnamefont {A.}~\bibnamefont {Fasolino}},
  \bibinfo {author} {\bibfnamefont {E.}~\bibnamefont {Molinari}}, \ and\
  \bibinfo {author} {\bibfnamefont {M.~C.}\ \bibnamefont {Righi}},\ }\href
  {\doibase 10.1103/PhysRevB.86.245434} {\bibfield  {journal} {\bibinfo
  {journal} {Phys. Rev. B}\ }\textbf {\bibinfo {volume} {86}},\ \bibinfo
  {pages} {245434} (\bibinfo {year} {2012})}\BibitemShut {NoStop}%
\bibitem [{\citenamefont {Lebedeva}\ \emph
  {et~al.}(2011{\natexlab{a}})\citenamefont {Lebedeva}, \citenamefont
  {Knizhnik}, \citenamefont {Popov}, \citenamefont {Ershova}, \citenamefont
  {Lozovik},\ and\ \citenamefont {Potapkin}}]{Lebedeva2011a}%
  \BibitemOpen
  \bibfield  {author} {\bibinfo {author} {\bibfnamefont {I.~V.}\ \bibnamefont
  {Lebedeva}}, \bibinfo {author} {\bibfnamefont {A.~A.}\ \bibnamefont
  {Knizhnik}}, \bibinfo {author} {\bibfnamefont {A.~M.}\ \bibnamefont {Popov}},
  \bibinfo {author} {\bibfnamefont {O.~V.}\ \bibnamefont {Ershova}}, \bibinfo
  {author} {\bibfnamefont {Y.~E.}\ \bibnamefont {Lozovik}}, \ and\ \bibinfo
  {author} {\bibfnamefont {B.~V.}\ \bibnamefont {Potapkin}},\ }\href {\doibase
  10.1063/1.3557819} {\bibfield  {journal} {\bibinfo  {journal} {J. Chem.
  Phys.}\ }\textbf {\bibinfo {volume} {134}},\ \bibinfo {pages} {104505}
  (\bibinfo {year} {2011}{\natexlab{a}})}\BibitemShut {NoStop}%
\bibitem [{\citenamefont {Savin}\ \emph {et~al.}(2015)\citenamefont {Savin},
  \citenamefont {Korznikova},\ and\ \citenamefont {Dmitriev}}]{Savin2015}%
  \BibitemOpen
  \bibfield  {author} {\bibinfo {author} {\bibfnamefont {A.~V.}\ \bibnamefont
  {Savin}}, \bibinfo {author} {\bibfnamefont {E.~A.}\ \bibnamefont
  {Korznikova}}, \ and\ \bibinfo {author} {\bibfnamefont {S.~V.}\ \bibnamefont
  {Dmitriev}},\ }\href {\doibase 10.1103/PhysRevB.92.035412} {\bibfield
  {journal} {\bibinfo  {journal} {Phys. Rev. B}\ }\textbf {\bibinfo {volume}
  {92}},\ \bibinfo {pages} {035412} (\bibinfo {year} {2015})}\BibitemShut
  {NoStop}%
\bibitem [{\citenamefont {Shi}\ \emph {et~al.}(2010{\natexlab{b}})\citenamefont
  {Shi}, \citenamefont {Pugno},\ and\ \citenamefont {Gao}}]{Shi2010a}%
  \BibitemOpen
  \bibfield  {author} {\bibinfo {author} {\bibfnamefont {X.}~\bibnamefont
  {Shi}}, \bibinfo {author} {\bibfnamefont {N.~M.}\ \bibnamefont {Pugno}}, \
  and\ \bibinfo {author} {\bibfnamefont {H.}~\bibnamefont {Gao}},\ }\href
  {\doibase 10.1166/jctn.2010.1387} {\bibfield  {journal} {\bibinfo  {journal}
  {J. Comput. Theor. Nanosci.}\ }\textbf {\bibinfo {volume} {7}},\ \bibinfo
  {pages} {517} (\bibinfo {year} {2010}{\natexlab{b}})}\BibitemShut {NoStop}%
\bibitem [{\citenamefont {Perim}\ and\ \citenamefont
  {Galv\~ao}(2009)}]{Perim2009}%
  \BibitemOpen
  \bibfield  {author} {\bibinfo {author} {\bibfnamefont {E.}~\bibnamefont
  {Perim}}\ and\ \bibinfo {author} {\bibfnamefont {D.~S.}\ \bibnamefont
  {Galv\~ao}},\ }\href {\doibase 10.1088/0957-4484/20/33/335702} {\bibfield
  {journal} {\bibinfo  {journal} {Nanotechnology}\ }\textbf {\bibinfo {volume}
  {20}},\ \bibinfo {pages} {335702} (\bibinfo {year} {2009})}\BibitemShut
  {NoStop}%
\bibitem [{\citenamefont {Arroyo}\ and\ \citenamefont
  {Belytschko}(2004)}]{Arroyo2004}%
  \BibitemOpen
  \bibfield  {author} {\bibinfo {author} {\bibfnamefont {M.}~\bibnamefont
  {Arroyo}}\ and\ \bibinfo {author} {\bibfnamefont {T.}~\bibnamefont
  {Belytschko}},\ }\href {\doibase 10.1103/PhysRevB.69.115415} {\bibfield
  {journal} {\bibinfo  {journal} {Phys. Rev. B}\ }\textbf {\bibinfo {volume}
  {69}},\ \bibinfo {pages} {115415} (\bibinfo {year} {2004})}\BibitemShut
  {NoStop}%
\bibitem [{\citenamefont {Kudin}\ \emph {et~al.}(2001)\citenamefont {Kudin},
  \citenamefont {Scuseria},\ and\ \citenamefont {Yakobson}}]{Kudin2001}%
  \BibitemOpen
  \bibfield  {author} {\bibinfo {author} {\bibfnamefont {K.~N.}\ \bibnamefont
  {Kudin}}, \bibinfo {author} {\bibfnamefont {G.~E.}\ \bibnamefont {Scuseria}},
  \ and\ \bibinfo {author} {\bibfnamefont {B.~I.}\ \bibnamefont {Yakobson}},\
  }\href {\doibase 10.1103/PhysRevB.64.235406} {\bibfield  {journal} {\bibinfo
  {journal} {Phys. Rev. B}\ }\textbf {\bibinfo {volume} {64}},\ \bibinfo
  {pages} {235406} (\bibinfo {year} {2001})}\BibitemShut {NoStop}%
\bibitem [{\citenamefont {Lebedeva}\ \emph {et~al.}(2012)\citenamefont
  {Lebedeva}, \citenamefont {Popov}, \citenamefont {Knizhnik}, \citenamefont
  {Khlobystov},\ and\ \citenamefont {Potapkin}}]{Lebedeva2012}%
  \BibitemOpen
  \bibfield  {author} {\bibinfo {author} {\bibfnamefont {I.~V.}\ \bibnamefont
  {Lebedeva}}, \bibinfo {author} {\bibfnamefont {A.~M.}\ \bibnamefont {Popov}},
  \bibinfo {author} {\bibfnamefont {A.~A.}\ \bibnamefont {Knizhnik}}, \bibinfo
  {author} {\bibfnamefont {A.~N.}\ \bibnamefont {Khlobystov}}, \ and\ \bibinfo
  {author} {\bibfnamefont {B.~V.}\ \bibnamefont {Potapkin}},\ }\href {\doibase
  10.1039/C2NR30144J} {\bibfield  {journal} {\bibinfo  {journal} {Nanoscale}\
  }\textbf {\bibinfo {volume} {4}},\ \bibinfo {pages} {4522} (\bibinfo {year}
  {2012})}\BibitemShut {NoStop}%
\bibitem [{\citenamefont {G\"ulseren}\ \emph {et~al.}(2002)\citenamefont
  {G\"ulseren}, \citenamefont {Yildirim},\ and\ \citenamefont
  {Ciraci}}]{Gulseren2002}%
  \BibitemOpen
  \bibfield  {author} {\bibinfo {author} {\bibfnamefont {O.}~\bibnamefont
  {G\"ulseren}}, \bibinfo {author} {\bibfnamefont {T.}~\bibnamefont
  {Yildirim}}, \ and\ \bibinfo {author} {\bibfnamefont {S.}~\bibnamefont
  {Ciraci}},\ }\href {\doibase 10.1103/PhysRevB.65.153405} {\bibfield
  {journal} {\bibinfo  {journal} {Phys. Rev. B}\ }\textbf {\bibinfo {volume}
  {65}},\ \bibinfo {pages} {153405} (\bibinfo {year} {2002})}\BibitemShut
  {NoStop}%
\bibitem [{\citenamefont {Cherian}\ and\ \citenamefont
  {Mahadevan}(2007)}]{Cherian2007}%
  \BibitemOpen
  \bibfield  {author} {\bibinfo {author} {\bibfnamefont {R.}~\bibnamefont
  {Cherian}}\ and\ \bibinfo {author} {\bibfnamefont {P.}~\bibnamefont
  {Mahadevan}},\ }\href {\doibase 10.1166/jnn.2007.714} {\bibfield  {journal}
  {\bibinfo  {journal} {J. Nanosci. Nanotechnol.}\ }\textbf {\bibinfo {volume}
  {7}},\ \bibinfo {pages} {1779} (\bibinfo {year} {2007})}\BibitemShut
  {NoStop}%
\bibitem [{\citenamefont {S\'anchez-Portal}\ \emph {et~al.}(1999)\citenamefont
  {S\'anchez-Portal}, \citenamefont {Artacho}, \citenamefont {Soler},
  \citenamefont {Rubio},\ and\ \citenamefont {Ordej\'on}}]{Sanchez1999}%
  \BibitemOpen
  \bibfield  {author} {\bibinfo {author} {\bibfnamefont {D.}~\bibnamefont
  {S\'anchez-Portal}}, \bibinfo {author} {\bibfnamefont {E.}~\bibnamefont
  {Artacho}}, \bibinfo {author} {\bibfnamefont {J.~M.}\ \bibnamefont {Soler}},
  \bibinfo {author} {\bibfnamefont {A.}~\bibnamefont {Rubio}}, \ and\ \bibinfo
  {author} {\bibfnamefont {P.}~\bibnamefont {Ordej\'on}},\ }\href {\doibase
  10.1103/PhysRevB.59.12678} {\bibfield  {journal} {\bibinfo  {journal} {Phys.
  Rev. B}\ }\textbf {\bibinfo {volume} {59}},\ \bibinfo {pages} {12678}
  (\bibinfo {year} {1999})}\BibitemShut {NoStop}%
\bibitem [{\citenamefont {K\"urti}\ \emph {et~al.}(1998)\citenamefont
  {K\"urti}, \citenamefont {Kresse},\ and\ \citenamefont
  {Kuzmany}}]{Kurti1998}%
  \BibitemOpen
  \bibfield  {author} {\bibinfo {author} {\bibfnamefont {J.}~\bibnamefont
  {K\"urti}}, \bibinfo {author} {\bibfnamefont {G.}~\bibnamefont {Kresse}}, \
  and\ \bibinfo {author} {\bibfnamefont {H.}~\bibnamefont {Kuzmany}},\ }\href
  {\doibase 10.1103/PhysRevB.58.R8869} {\bibfield  {journal} {\bibinfo
  {journal} {Phys. Rev. B}\ }\textbf {\bibinfo {volume} {58}},\ \bibinfo
  {pages} {R8869} (\bibinfo {year} {1998})}\BibitemShut {NoStop}%
\bibitem [{\citenamefont {Fogler}\ \emph {et~al.}(2010)\citenamefont {Fogler},
  \citenamefont {Castro~Neto},\ and\ \citenamefont {Guinea}}]{Fogler2010}%
  \BibitemOpen
  \bibfield  {author} {\bibinfo {author} {\bibfnamefont {M.~M.}\ \bibnamefont
  {Fogler}}, \bibinfo {author} {\bibfnamefont {A.~H.}\ \bibnamefont
  {Castro~Neto}}, \ and\ \bibinfo {author} {\bibfnamefont {F.}~\bibnamefont
  {Guinea}},\ }\href {\doibase 10.1103/PhysRevB.81.161408} {\bibfield
  {journal} {\bibinfo  {journal} {Phys. Rev. B}\ }\textbf {\bibinfo {volume}
  {81}},\ \bibinfo {pages} {161408} (\bibinfo {year} {2010})}\BibitemShut
  {NoStop}%
\bibitem [{\citenamefont {Shi}\ \emph {et~al.}(2011)\citenamefont {Shi},
  \citenamefont {Pugno},\ and\ \citenamefont {Gao}}]{Shi2011}%
  \BibitemOpen
  \bibfield  {author} {\bibinfo {author} {\bibfnamefont {X.}~\bibnamefont
  {Shi}}, \bibinfo {author} {\bibfnamefont {N.~M.}\ \bibnamefont {Pugno}}, \
  and\ \bibinfo {author} {\bibfnamefont {H.}~\bibnamefont {Gao}},\ }\href
  {\doibase 10.1007/s10704-010-9545-y} {\bibfield  {journal} {\bibinfo
  {journal} {Int. J. Fract.}\ }\textbf {\bibinfo {volume} {171}},\ \bibinfo
  {pages} {163} (\bibinfo {year} {2011})}\BibitemShut {NoStop}%
\bibitem [{\citenamefont {Siahlo}\ \emph {et~al.}(2017)\citenamefont {Siahlo},
  \citenamefont {Popov}, \citenamefont {Poklonski},\ and\ \citenamefont
  {Lozovik}}]{Siahlo2017}%
  \BibitemOpen
  \bibfield  {author} {\bibinfo {author} {\bibfnamefont {A.~I.}\ \bibnamefont
  {Siahlo}}, \bibinfo {author} {\bibfnamefont {A.~M.}\ \bibnamefont {Popov}},
  \bibinfo {author} {\bibfnamefont {N.~A.}\ \bibnamefont {Poklonski}}, \ and\
  \bibinfo {author} {\bibfnamefont {Y.~E.}\ \bibnamefont {Lozovik}},\ }\href
  {\doibase 10.1134/S1063785017070264} {\bibfield  {journal} {\bibinfo
  {journal} {Tech. Phys. Lett.}\ }\textbf {\bibinfo {volume} {43}},\ \bibinfo
  {pages} {662} (\bibinfo {year} {2017})}\BibitemShut {NoStop}%
\bibitem [{\citenamefont {Yamaletdinov}\ and\ \citenamefont
  {Pershin}(2017)}]{Yamaletdinov2017}%
  \BibitemOpen
  \bibfield  {author} {\bibinfo {author} {\bibfnamefont {R.~D.}\ \bibnamefont
  {Yamaletdinov}}\ and\ \bibinfo {author} {\bibfnamefont {Y.~V.}\ \bibnamefont
  {Pershin}},\ }\href {\doibase 10.1038/srep42356} {\bibfield  {journal}
  {\bibinfo  {journal} {Sci. Rep.}\ }\textbf {\bibinfo {volume} {7}},\ \bibinfo
  {pages} {42356} (\bibinfo {year} {2017})}\BibitemShut {NoStop}%
\bibitem [{\citenamefont {Gao}\ \emph {et~al.}(2010)\citenamefont {Gao},
  \citenamefont {Chen}, \citenamefont {Xu}, \citenamefont {Zou}, \citenamefont
  {Gu}, \citenamefont {Xu}, \citenamefont {Jen},\ and\ \citenamefont
  {Chen}}]{Gao2010}%
  \BibitemOpen
  \bibfield  {author} {\bibinfo {author} {\bibfnamefont {Y.}~\bibnamefont
  {Gao}}, \bibinfo {author} {\bibfnamefont {X.}~\bibnamefont {Chen}}, \bibinfo
  {author} {\bibfnamefont {H.}~\bibnamefont {Xu}}, \bibinfo {author}
  {\bibfnamefont {Y.}~\bibnamefont {Zou}}, \bibinfo {author} {\bibfnamefont
  {R.}~\bibnamefont {Gu}}, \bibinfo {author} {\bibfnamefont {M.}~\bibnamefont
  {Xu}}, \bibinfo {author} {\bibfnamefont {A.~K.-Y.}\ \bibnamefont {Jen}}, \
  and\ \bibinfo {author} {\bibfnamefont {H.}~\bibnamefont {Chen}},\ }\href
  {\doibase 10.1016/j.carbon.2010.08.007} {\bibfield  {journal} {\bibinfo
  {journal} {Carbon}\ }\textbf {\bibinfo {volume} {48}},\ \bibinfo {pages}
  {4475} (\bibinfo {year} {2010})}\BibitemShut {NoStop}%
\bibitem [{\citenamefont {Kim}\ and\ \citenamefont {Min}(2010)}]{Kim2010}%
  \BibitemOpen
  \bibfield  {author} {\bibinfo {author} {\bibfnamefont {Y.-K.}\ \bibnamefont
  {Kim}}\ and\ \bibinfo {author} {\bibfnamefont {D.-H.}\ \bibnamefont {Min}},\
  }\href {\doibase https://doi.org/10.1016/j.carbon.2010.07.039} {\bibfield
  {journal} {\bibinfo  {journal} {Carbon}\ }\textbf {\bibinfo {volume} {48}},\
  \bibinfo {pages} {4283} (\bibinfo {year} {2010})}\BibitemShut {NoStop}%
\bibitem [{\citenamefont {Wang}\ \emph {et~al.}(2012)\citenamefont {Wang},
  \citenamefont {Yang}, \citenamefont {Huang}, \citenamefont {Huang},
  \citenamefont {Shen}, \citenamefont {Guo}, \citenamefont {Mei},\ and\
  \citenamefont {Cui}}]{Wang2012}%
  \BibitemOpen
  \bibfield  {author} {\bibinfo {author} {\bibfnamefont {X.}~\bibnamefont
  {Wang}}, \bibinfo {author} {\bibfnamefont {D.-P.}\ \bibnamefont {Yang}},
  \bibinfo {author} {\bibfnamefont {G.}~\bibnamefont {Huang}}, \bibinfo
  {author} {\bibfnamefont {P.}~\bibnamefont {Huang}}, \bibinfo {author}
  {\bibfnamefont {G.}~\bibnamefont {Shen}}, \bibinfo {author} {\bibfnamefont
  {S.}~\bibnamefont {Guo}}, \bibinfo {author} {\bibfnamefont {Y.}~\bibnamefont
  {Mei}}, \ and\ \bibinfo {author} {\bibfnamefont {D.}~\bibnamefont {Cui}},\
  }\href {\doibase 10.1039/C2JM32810K} {\bibfield  {journal} {\bibinfo
  {journal} {J. Mater. Chem.}\ }\textbf {\bibinfo {volume} {22}},\ \bibinfo
  {pages} {17441} (\bibinfo {year} {2012})}\BibitemShut {NoStop}%
\bibitem [{\citenamefont {Chen}\ \emph {et~al.}(2013)\citenamefont {Chen},
  \citenamefont {Boulos}, \citenamefont {Dobson},\ and\ \citenamefont
  {Raston}}]{Chen2012}%
  \BibitemOpen
  \bibfield  {author} {\bibinfo {author} {\bibfnamefont {X.}~\bibnamefont
  {Chen}}, \bibinfo {author} {\bibfnamefont {R.~A.}\ \bibnamefont {Boulos}},
  \bibinfo {author} {\bibfnamefont {J.~F.}\ \bibnamefont {Dobson}}, \ and\
  \bibinfo {author} {\bibfnamefont {C.~L.}\ \bibnamefont {Raston}},\ }\href
  {\doibase 10.1039/C2NR33071G} {\bibfield  {journal} {\bibinfo  {journal}
  {Nanoscale}\ }\textbf {\bibinfo {volume} {5}},\ \bibinfo {pages} {498}
  (\bibinfo {year} {2013})}\BibitemShut {NoStop}%
\bibitem [{\citenamefont {Li}\ \emph {et~al.}(2013)\citenamefont {Li},
  \citenamefont {Hao}, \citenamefont {Zhao}, \citenamefont {Wu}, \citenamefont
  {Yang}, \citenamefont {Tian},\ and\ \citenamefont {Qian}}]{Li2013}%
  \BibitemOpen
  \bibfield  {author} {\bibinfo {author} {\bibfnamefont {X.}~\bibnamefont
  {Li}}, \bibinfo {author} {\bibfnamefont {X.}~\bibnamefont {Hao}}, \bibinfo
  {author} {\bibfnamefont {M.}~\bibnamefont {Zhao}}, \bibinfo {author}
  {\bibfnamefont {Y.}~\bibnamefont {Wu}}, \bibinfo {author} {\bibfnamefont
  {J.}~\bibnamefont {Yang}}, \bibinfo {author} {\bibfnamefont {Y.}~\bibnamefont
  {Tian}}, \ and\ \bibinfo {author} {\bibfnamefont {G.}~\bibnamefont {Qian}},\
  }\href {\doibase 10.1002/adma.201204031} {\bibfield  {journal} {\bibinfo
  {journal} {Adv. Mater.}\ }\textbf {\bibinfo {volume} {25}},\ \bibinfo {pages}
  {2200} (\bibinfo {year} {2013})}\BibitemShut {NoStop}%
\bibitem [{\citenamefont {Hwang}\ and\ \citenamefont {Suh}(2014)}]{Hwang2014}%
  \BibitemOpen
  \bibfield  {author} {\bibinfo {author} {\bibfnamefont {D.~Y.}\ \bibnamefont
  {Hwang}}\ and\ \bibinfo {author} {\bibfnamefont {D.~H.}\ \bibnamefont
  {Suh}},\ }\href {\doibase 10.1039/C4NR00897A} {\bibfield  {journal} {\bibinfo
   {journal} {Nanoscale}\ }\textbf {\bibinfo {volume} {6}},\ \bibinfo {pages}
  {5686} (\bibinfo {year} {2014})}\BibitemShut {NoStop}%
\bibitem [{\citenamefont {Wang}\ \emph {et~al.}(2017)\citenamefont {Wang},
  \citenamefont {Ma},\ and\ \citenamefont {Sun}}]{Wang2017}%
  \BibitemOpen
  \bibfield  {author} {\bibinfo {author} {\bibfnamefont {J.}~\bibnamefont
  {Wang}}, \bibinfo {author} {\bibfnamefont {F.}~\bibnamefont {Ma}}, \ and\
  \bibinfo {author} {\bibfnamefont {M.}~\bibnamefont {Sun}},\ }\href {\doibase
  10.1039/C7RA00260B} {\bibfield  {journal} {\bibinfo  {journal} {RSC Adv.}\
  }\textbf {\bibinfo {volume} {7}},\ \bibinfo {pages} {16801} (\bibinfo {year}
  {2017})}\BibitemShut {NoStop}%
\bibitem [{\citenamefont {Li}\ \emph {et~al.}(2016)\citenamefont {Li},
  \citenamefont {Liu}, \citenamefont {Zhang},\ and\ \citenamefont
  {Liu}}]{Li2016}%
  \BibitemOpen
  \bibfield  {author} {\bibinfo {author} {\bibfnamefont {Q.}~\bibnamefont
  {Li}}, \bibinfo {author} {\bibfnamefont {M.}~\bibnamefont {Liu}}, \bibinfo
  {author} {\bibfnamefont {Y.}~\bibnamefont {Zhang}}, \ and\ \bibinfo {author}
  {\bibfnamefont {Z.}~\bibnamefont {Liu}},\ }\href {\doibase
  10.1002/smll.201501766} {\bibfield  {journal} {\bibinfo  {journal} {Small}\
  }\textbf {\bibinfo {volume} {12}},\ \bibinfo {pages} {32} (\bibinfo {year}
  {2016})}\BibitemShut {NoStop}%
\bibitem [{\citenamefont {Ramakrishna}(2005)}]{Ramakrishna2005}%
  \BibitemOpen
  \bibfield  {author} {\bibinfo {author} {\bibfnamefont {S.~A.}\ \bibnamefont
  {Ramakrishna}},\ }\href {\doibase 10.1088/0034-4885/68/2/R06} {\bibfield
  {journal} {\bibinfo  {journal} {Rep. Prog. Phys.}\ }\textbf {\bibinfo
  {volume} {68}},\ \bibinfo {pages} {449} (\bibinfo {year} {2005})}\BibitemShut
  {NoStop}%
\bibitem [{\citenamefont {Zacharia}\ \emph {et~al.}(2004)\citenamefont
  {Zacharia}, \citenamefont {Ulbricht},\ and\ \citenamefont
  {Hertel}}]{Zacharia2004}%
  \BibitemOpen
  \bibfield  {author} {\bibinfo {author} {\bibfnamefont {R.}~\bibnamefont
  {Zacharia}}, \bibinfo {author} {\bibfnamefont {H.}~\bibnamefont {Ulbricht}},
  \ and\ \bibinfo {author} {\bibfnamefont {T.}~\bibnamefont {Hertel}},\ }\href
  {\doibase 10.1103/PhysRevB.69.155406} {\bibfield  {journal} {\bibinfo
  {journal} {Phys. Rev. B}\ }\textbf {\bibinfo {volume} {69}},\ \bibinfo
  {pages} {155406} (\bibinfo {year} {2004})}\BibitemShut {NoStop}%
\bibitem [{\citenamefont {Girifalco}\ and\ \citenamefont
  {Lad}(1956)}]{Girifalco1956}%
  \BibitemOpen
  \bibfield  {author} {\bibinfo {author} {\bibfnamefont {L.~A.}\ \bibnamefont
  {Girifalco}}\ and\ \bibinfo {author} {\bibfnamefont {R.~A.}\ \bibnamefont
  {Lad}},\ }\href {\doibase 10.1063/1.1743030} {\bibfield  {journal} {\bibinfo
  {journal} {J. Chem. Phys.}\ }\textbf {\bibinfo {volume} {25}},\ \bibinfo
  {pages} {693} (\bibinfo {year} {1956})}\BibitemShut {NoStop}%
\bibitem [{\citenamefont {Benedict}\ \emph {et~al.}(1998)\citenamefont
  {Benedict}, \citenamefont {Chopra}, \citenamefont {Cohen}, \citenamefont
  {Zettl}, \citenamefont {Louie},\ and\ \citenamefont {Crespi}}]{Benedict1998}%
  \BibitemOpen
  \bibfield  {author} {\bibinfo {author} {\bibfnamefont {L.~X.}\ \bibnamefont
  {Benedict}}, \bibinfo {author} {\bibfnamefont {N.~G.}\ \bibnamefont
  {Chopra}}, \bibinfo {author} {\bibfnamefont {M.~L.}\ \bibnamefont {Cohen}},
  \bibinfo {author} {\bibfnamefont {A.}~\bibnamefont {Zettl}}, \bibinfo
  {author} {\bibfnamefont {S.~G.}\ \bibnamefont {Louie}}, \ and\ \bibinfo
  {author} {\bibfnamefont {V.~H.}\ \bibnamefont {Crespi}},\ }\href {\doibase
  https://doi.org/10.1016/S0009-2614(97)01466-8} {\bibfield  {journal}
  {\bibinfo  {journal} {Chem. Phys. Lett.}\ }\textbf {\bibinfo {volume}
  {286}},\ \bibinfo {pages} {490} (\bibinfo {year} {1998})}\BibitemShut
  {NoStop}%
\bibitem [{\citenamefont {Liu}\ \emph {et~al.}(2012)\citenamefont {Liu},
  \citenamefont {Liu}, \citenamefont {Cheng}, \citenamefont {Li}, \citenamefont
  {Wang},\ and\ \citenamefont {Zheng}}]{Liu2012}%
  \BibitemOpen
  \bibfield  {author} {\bibinfo {author} {\bibfnamefont {Z.}~\bibnamefont
  {Liu}}, \bibinfo {author} {\bibfnamefont {J.~Z.}\ \bibnamefont {Liu}},
  \bibinfo {author} {\bibfnamefont {Y.}~\bibnamefont {Cheng}}, \bibinfo
  {author} {\bibfnamefont {Z.}~\bibnamefont {Li}}, \bibinfo {author}
  {\bibfnamefont {L.}~\bibnamefont {Wang}}, \ and\ \bibinfo {author}
  {\bibfnamefont {Q.}~\bibnamefont {Zheng}},\ }\href {\doibase
  10.1103/PhysRevB.85.205418} {\bibfield  {journal} {\bibinfo  {journal} {Phys.
  Rev. B}\ }\textbf {\bibinfo {volume} {85}},\ \bibinfo {pages} {205418}
  (\bibinfo {year} {2012})}\BibitemShut {NoStop}%
\bibitem [{\citenamefont {Leb\`egue}\ \emph {et~al.}(2010)\citenamefont
  {Leb\`egue}, \citenamefont {Harl}, \citenamefont {Gould}, \citenamefont
  {\'Angy\'an}, \citenamefont {Kresse},\ and\ \citenamefont
  {Dobson}}]{Lebegue2010}%
  \BibitemOpen
  \bibfield  {author} {\bibinfo {author} {\bibfnamefont {S.}~\bibnamefont
  {Leb\`egue}}, \bibinfo {author} {\bibfnamefont {J.}~\bibnamefont {Harl}},
  \bibinfo {author} {\bibfnamefont {T.}~\bibnamefont {Gould}}, \bibinfo
  {author} {\bibfnamefont {J.~G.}\ \bibnamefont {\'Angy\'an}}, \bibinfo
  {author} {\bibfnamefont {G.}~\bibnamefont {Kresse}}, \ and\ \bibinfo {author}
  {\bibfnamefont {J.~F.}\ \bibnamefont {Dobson}},\ }\href {\doibase
  10.1103/PhysRevLett.105.196401} {\bibfield  {journal} {\bibinfo  {journal}
  {Phys. Rev. Lett.}\ }\textbf {\bibinfo {volume} {105}},\ \bibinfo {pages}
  {196401} (\bibinfo {year} {2010})}\BibitemShut {NoStop}%
\bibitem [{\citenamefont {Spanu}\ \emph {et~al.}(2009)\citenamefont {Spanu},
  \citenamefont {Sorella},\ and\ \citenamefont {Galli}}]{Spanu2009}%
  \BibitemOpen
  \bibfield  {author} {\bibinfo {author} {\bibfnamefont {L.}~\bibnamefont
  {Spanu}}, \bibinfo {author} {\bibfnamefont {S.}~\bibnamefont {Sorella}}, \
  and\ \bibinfo {author} {\bibfnamefont {G.}~\bibnamefont {Galli}},\ }\href
  {\doibase 10.1103/PhysRevLett.103.196401} {\bibfield  {journal} {\bibinfo
  {journal} {Phys. Rev. Lett.}\ }\textbf {\bibinfo {volume} {103}},\ \bibinfo
  {pages} {196401} (\bibinfo {year} {2009})}\BibitemShut {NoStop}%
\bibitem [{\citenamefont {Mostaani}\ \emph {et~al.}(2015)\citenamefont
  {Mostaani}, \citenamefont {Drummond},\ and\ \citenamefont
  {Fal'ko}}]{Mostaani2015}%
  \BibitemOpen
  \bibfield  {author} {\bibinfo {author} {\bibfnamefont {E.}~\bibnamefont
  {Mostaani}}, \bibinfo {author} {\bibfnamefont {N.~D.}\ \bibnamefont
  {Drummond}}, \ and\ \bibinfo {author} {\bibfnamefont {V.~I.}\ \bibnamefont
  {Fal'ko}},\ }\href {\doibase 10.1103/PhysRevLett.115.115501} {\bibfield
  {journal} {\bibinfo  {journal} {Phys. Rev. Lett.}\ }\textbf {\bibinfo
  {volume} {115}},\ \bibinfo {pages} {115501} (\bibinfo {year}
  {2015})}\BibitemShut {NoStop}%
\bibitem [{\citenamefont {Zhou}\ \emph {et~al.}(2015)\citenamefont {Zhou},
  \citenamefont {Han}, \citenamefont {Dai}, \citenamefont {Sun},\ and\
  \citenamefont {Srolovitz}}]{Zhou2015}%
  \BibitemOpen
  \bibfield  {author} {\bibinfo {author} {\bibfnamefont {S.}~\bibnamefont
  {Zhou}}, \bibinfo {author} {\bibfnamefont {J.}~\bibnamefont {Han}}, \bibinfo
  {author} {\bibfnamefont {S.}~\bibnamefont {Dai}}, \bibinfo {author}
  {\bibfnamefont {J.}~\bibnamefont {Sun}}, \ and\ \bibinfo {author}
  {\bibfnamefont {D.~J.}\ \bibnamefont {Srolovitz}},\ }\href {\doibase
  10.1103/PhysRevB.92.155438} {\bibfield  {journal} {\bibinfo  {journal} {Phys.
  Rev. B}\ }\textbf {\bibinfo {volume} {92}},\ \bibinfo {pages} {155438}
  (\bibinfo {year} {2015})}\BibitemShut {NoStop}%
\bibitem [{\citenamefont {Lebedeva}\ \emph
  {et~al.}(2011{\natexlab{b}})\citenamefont {Lebedeva}, \citenamefont
  {Knizhnik}, \citenamefont {Popov}, \citenamefont {Lozovik},\ and\
  \citenamefont {Potapkin}}]{Lebedeva2011}%
  \BibitemOpen
  \bibfield  {author} {\bibinfo {author} {\bibfnamefont {I.~V.}\ \bibnamefont
  {Lebedeva}}, \bibinfo {author} {\bibfnamefont {A.~A.}\ \bibnamefont
  {Knizhnik}}, \bibinfo {author} {\bibfnamefont {A.~M.}\ \bibnamefont {Popov}},
  \bibinfo {author} {\bibfnamefont {Y.~E.}\ \bibnamefont {Lozovik}}, \ and\
  \bibinfo {author} {\bibfnamefont {B.~V.}\ \bibnamefont {Potapkin}},\ }\href
  {\doibase 10.1039/C0CP02614J} {\bibfield  {journal} {\bibinfo  {journal}
  {Phys. Chem. Chem. Phys.}\ }\textbf {\bibinfo {volume} {13}},\ \bibinfo
  {pages} {5687} (\bibinfo {year} {2011}{\natexlab{b}})}\BibitemShut {NoStop}%
\bibitem [{\citenamefont {Sachs}\ \emph {et~al.}(2011)\citenamefont {Sachs},
  \citenamefont {Wehling}, \citenamefont {Katsnelson},\ and\ \citenamefont
  {Lichtenstein}}]{Sachs2011}%
  \BibitemOpen
  \bibfield  {author} {\bibinfo {author} {\bibfnamefont {B.}~\bibnamefont
  {Sachs}}, \bibinfo {author} {\bibfnamefont {T.~O.}\ \bibnamefont {Wehling}},
  \bibinfo {author} {\bibfnamefont {M.~I.}\ \bibnamefont {Katsnelson}}, \ and\
  \bibinfo {author} {\bibfnamefont {A.~I.}\ \bibnamefont {Lichtenstein}},\
  }\href {\doibase 10.1103/PhysRevB.84.195414} {\bibfield  {journal} {\bibinfo
  {journal} {Phys. Rev. B}\ }\textbf {\bibinfo {volume} {84}},\ \bibinfo
  {pages} {195414} (\bibinfo {year} {2011})}\BibitemShut {NoStop}%
\bibitem [{\citenamefont {Popov}\ \emph {et~al.}(2012)\citenamefont {Popov},
  \citenamefont {Lebedeva}, \citenamefont {Knizhnik}, \citenamefont {Lozovik},\
  and\ \citenamefont {Potapkin}}]{Popov2012}%
  \BibitemOpen
  \bibfield  {author} {\bibinfo {author} {\bibfnamefont {A.~M.}\ \bibnamefont
  {Popov}}, \bibinfo {author} {\bibfnamefont {I.~V.}\ \bibnamefont {Lebedeva}},
  \bibinfo {author} {\bibfnamefont {A.~A.}\ \bibnamefont {Knizhnik}}, \bibinfo
  {author} {\bibfnamefont {Y.~E.}\ \bibnamefont {Lozovik}}, \ and\ \bibinfo
  {author} {\bibfnamefont {B.~V.}\ \bibnamefont {Potapkin}},\ }\href {\doibase
  10.1016/j.cplett.2012.03.082} {\bibfield  {journal} {\bibinfo  {journal}
  {Chem. Phys. Lett.}\ }\textbf {\bibinfo {volume} {536}},\ \bibinfo {pages}
  {82} (\bibinfo {year} {2012})}\BibitemShut {NoStop}%
\bibitem [{\citenamefont {Lebedev}\ \emph {et~al.}(2016)\citenamefont
  {Lebedev}, \citenamefont {Lebedeva}, \citenamefont {Knizhnik},\ and\
  \citenamefont {Popov}}]{Lebedev2016}%
  \BibitemOpen
  \bibfield  {author} {\bibinfo {author} {\bibfnamefont {A.~V.}\ \bibnamefont
  {Lebedev}}, \bibinfo {author} {\bibfnamefont {I.~V.}\ \bibnamefont
  {Lebedeva}}, \bibinfo {author} {\bibfnamefont {A.~A.}\ \bibnamefont
  {Knizhnik}}, \ and\ \bibinfo {author} {\bibfnamefont {A.~M.}\ \bibnamefont
  {Popov}},\ }\href {\doibase 10.1039/C5RA20882C} {\bibfield  {journal}
  {\bibinfo  {journal} {RSC Adv.}\ }\textbf {\bibinfo {volume} {6}},\ \bibinfo
  {pages} {6423} (\bibinfo {year} {2016})}\BibitemShut {NoStop}%
\bibitem [{\citenamefont {Lebedeva}\ \emph {et~al.}(2017)\citenamefont
  {Lebedeva}, \citenamefont {Lebedev}, \citenamefont {Popov},\ and\
  \citenamefont {Knizhnik}}]{Lebedeva2017}%
  \BibitemOpen
  \bibfield  {author} {\bibinfo {author} {\bibfnamefont {I.~V.}\ \bibnamefont
  {Lebedeva}}, \bibinfo {author} {\bibfnamefont {A.~V.}\ \bibnamefont
  {Lebedev}}, \bibinfo {author} {\bibfnamefont {A.~M.}\ \bibnamefont {Popov}},
  \ and\ \bibinfo {author} {\bibfnamefont {A.~A.}\ \bibnamefont {Knizhnik}},\
  }\href {\doibase 10.1016/j.commatsci.2016.11.011} {\bibfield  {journal}
  {\bibinfo  {journal} {Comput. Mater. Sci.}\ }\textbf {\bibinfo {volume}
  {128}},\ \bibinfo {pages} {45} (\bibinfo {year} {2017})}\BibitemShut
  {NoStop}%
\bibitem [{\citenamefont {Perdew}\ \emph {et~al.}(1996)\citenamefont {Perdew},
  \citenamefont {Burke},\ and\ \citenamefont {Ernzerhof}}]{Perdew1996}%
  \BibitemOpen
  \bibfield  {author} {\bibinfo {author} {\bibfnamefont {J.~P.}\ \bibnamefont
  {Perdew}}, \bibinfo {author} {\bibfnamefont {K.}~\bibnamefont {Burke}}, \
  and\ \bibinfo {author} {\bibfnamefont {M.}~\bibnamefont {Ernzerhof}},\ }\href
  {\doibase 10.1103/PhysRevLett.77.3865} {\bibfield  {journal} {\bibinfo
  {journal} {Phys. Rev. Lett.}\ }\textbf {\bibinfo {volume} {77}},\ \bibinfo
  {pages} {3865} (\bibinfo {year} {1996})}\BibitemShut {NoStop}%
\bibitem [{\citenamefont {Lebedev}\ \emph {et~al.}(2017)\citenamefont
  {Lebedev}, \citenamefont {Lebedeva}, \citenamefont {Popov},\ and\
  \citenamefont {Knizhnik}}]{Lebedev2017}%
  \BibitemOpen
  \bibfield  {author} {\bibinfo {author} {\bibfnamefont {A.~V.}\ \bibnamefont
  {Lebedev}}, \bibinfo {author} {\bibfnamefont {I.~V.}\ \bibnamefont
  {Lebedeva}}, \bibinfo {author} {\bibfnamefont {A.~M.}\ \bibnamefont {Popov}},
  \ and\ \bibinfo {author} {\bibfnamefont {A.~A.}\ \bibnamefont {Knizhnik}},\
  }\href {\doibase 10.1103/PhysRevB.96.085432} {\bibfield  {journal} {\bibinfo
  {journal} {Phys. Rev. B}\ }\textbf {\bibinfo {volume} {96}},\ \bibinfo
  {pages} {085432} (\bibinfo {year} {2017})}\BibitemShut {NoStop}%
\bibitem [{\citenamefont {Lebedeva}\ \emph {et~al.}(2016)\citenamefont
  {Lebedeva}, \citenamefont {Lebedev}, \citenamefont {Popov},\ and\
  \citenamefont {Knizhnik}}]{Lebedeva2016}%
  \BibitemOpen
  \bibfield  {author} {\bibinfo {author} {\bibfnamefont {I.~V.}\ \bibnamefont
  {Lebedeva}}, \bibinfo {author} {\bibfnamefont {A.~V.}\ \bibnamefont
  {Lebedev}}, \bibinfo {author} {\bibfnamefont {A.~M.}\ \bibnamefont {Popov}},
  \ and\ \bibinfo {author} {\bibfnamefont {A.~A.}\ \bibnamefont {Knizhnik}},\
  }\href {\doibase 10.1103/PhysRevB.93.235414} {\bibfield  {journal} {\bibinfo
  {journal} {Phys. Rev. B}\ }\textbf {\bibinfo {volume} {93}},\ \bibinfo
  {pages} {235414} (\bibinfo {year} {2016})}\BibitemShut {NoStop}%
\bibitem [{\citenamefont {Constantinescu}\ \emph {et~al.}(2013)\citenamefont
  {Constantinescu}, \citenamefont {Kuc},\ and\ \citenamefont
  {Heine}}]{Constantinescu2013}%
  \BibitemOpen
  \bibfield  {author} {\bibinfo {author} {\bibfnamefont {G.}~\bibnamefont
  {Constantinescu}}, \bibinfo {author} {\bibfnamefont {A.}~\bibnamefont {Kuc}},
  \ and\ \bibinfo {author} {\bibfnamefont {T.}~\bibnamefont {Heine}},\ }\href
  {\doibase 10.1103/PhysRevLett.111.036104} {\bibfield  {journal} {\bibinfo
  {journal} {Phys. Rev. Lett.}\ }\textbf {\bibinfo {volume} {111}},\ \bibinfo
  {pages} {036104} (\bibinfo {year} {2013})}\BibitemShut {NoStop}%
\bibitem [{\citenamefont {Hern{\'a}ndez}\ \emph {et~al.}(1999)\citenamefont
  {Hern{\'a}ndez}, \citenamefont {Goze}, \citenamefont {Bernier},\ and\
  \citenamefont {Rubio}}]{Hernandez1999}%
  \BibitemOpen
  \bibfield  {author} {\bibinfo {author} {\bibfnamefont {E.}~\bibnamefont
  {Hern{\'a}ndez}}, \bibinfo {author} {\bibfnamefont {C.}~\bibnamefont {Goze}},
  \bibinfo {author} {\bibfnamefont {P.}~\bibnamefont {Bernier}}, \ and\
  \bibinfo {author} {\bibfnamefont {A.}~\bibnamefont {Rubio}},\ }\href
  {\doibase 10.1007/s003390050890} {\bibfield  {journal} {\bibinfo  {journal}
  {Appl. Phys. A}\ }\textbf {\bibinfo {volume} {68}},\ \bibinfo {pages} {287}
  (\bibinfo {year} {1999})}\BibitemShut {NoStop}%
\bibitem [{\citenamefont {Hern\'andez}\ \emph {et~al.}(1998)\citenamefont
  {Hern\'andez}, \citenamefont {Goze}, \citenamefont {Bernier},\ and\
  \citenamefont {Rubio}}]{Hernandez1998}%
  \BibitemOpen
  \bibfield  {author} {\bibinfo {author} {\bibfnamefont {E.}~\bibnamefont
  {Hern\'andez}}, \bibinfo {author} {\bibfnamefont {C.}~\bibnamefont {Goze}},
  \bibinfo {author} {\bibfnamefont {P.}~\bibnamefont {Bernier}}, \ and\
  \bibinfo {author} {\bibfnamefont {A.}~\bibnamefont {Rubio}},\ }\href
  {\doibase 10.1103/PhysRevLett.80.4502} {\bibfield  {journal} {\bibinfo
  {journal} {Phys. Rev. Lett.}\ }\textbf {\bibinfo {volume} {80}},\ \bibinfo
  {pages} {4502} (\bibinfo {year} {1998})}\BibitemShut {NoStop}%
\bibitem [{\citenamefont {Blase}\ \emph {et~al.}(1994)\citenamefont {Blase},
  \citenamefont {Rubio}, \citenamefont {Louie},\ and\ \citenamefont
  {Cohen}}]{Blase1994}%
  \BibitemOpen
  \bibfield  {author} {\bibinfo {author} {\bibfnamefont {X.}~\bibnamefont
  {Blase}}, \bibinfo {author} {\bibfnamefont {A.}~\bibnamefont {Rubio}},
  \bibinfo {author} {\bibfnamefont {S.~G.}\ \bibnamefont {Louie}}, \ and\
  \bibinfo {author} {\bibfnamefont {M.~L.}\ \bibnamefont {Cohen}},\ }\href
  {\doibase 10.1209/0295-5075/28/5/007} {\bibfield  {journal} {\bibinfo
  {journal} {Europhys. Lett.}\ }\textbf {\bibinfo {volume} {28}},\ \bibinfo
  {pages} {335} (\bibinfo {year} {1994})}\BibitemShut {NoStop}%
\bibitem [{\citenamefont {Xiang}\ \emph {et~al.}(2003)\citenamefont {Xiang},
  \citenamefont {Yang}, \citenamefont {Hou},\ and\ \citenamefont
  {Zhu}}]{Xiang2003}%
  \BibitemOpen
  \bibfield  {author} {\bibinfo {author} {\bibfnamefont {H.~J.}\ \bibnamefont
  {Xiang}}, \bibinfo {author} {\bibfnamefont {J.}~\bibnamefont {Yang}},
  \bibinfo {author} {\bibfnamefont {J.~G.}\ \bibnamefont {Hou}}, \ and\
  \bibinfo {author} {\bibfnamefont {Q.}~\bibnamefont {Zhu}},\ }\href {\doibase
  10.1103/PhysRevB.68.035427} {\bibfield  {journal} {\bibinfo  {journal} {Phys.
  Rev. B}\ }\textbf {\bibinfo {volume} {68}},\ \bibinfo {pages} {035427}
  (\bibinfo {year} {2003})}\BibitemShut {NoStop}%
\bibitem [{\citenamefont {Baumeier}\ \emph {et~al.}(2007)\citenamefont
  {Baumeier}, \citenamefont {Kr\"uger},\ and\ \citenamefont
  {Pollmann}}]{Baumeier2007}%
  \BibitemOpen
  \bibfield  {author} {\bibinfo {author} {\bibfnamefont {B.}~\bibnamefont
  {Baumeier}}, \bibinfo {author} {\bibfnamefont {P.}~\bibnamefont {Kr\"uger}},
  \ and\ \bibinfo {author} {\bibfnamefont {J.}~\bibnamefont {Pollmann}},\
  }\href {\doibase 10.1103/PhysRevB.76.085407} {\bibfield  {journal} {\bibinfo
  {journal} {Phys. Rev. B}\ }\textbf {\bibinfo {volume} {76}},\ \bibinfo
  {pages} {085407} (\bibinfo {year} {2007})}\BibitemShut {NoStop}%
\bibitem [{\citenamefont {Kresse}\ and\ \citenamefont
  {Furthm\"uller}(1996)}]{Kresse1996}%
  \BibitemOpen
  \bibfield  {author} {\bibinfo {author} {\bibfnamefont {G.}~\bibnamefont
  {Kresse}}\ and\ \bibinfo {author} {\bibfnamefont {J.}~\bibnamefont
  {Furthm\"uller}},\ }\href {\doibase 10.1103/PhysRevB.54.11169} {\bibfield
  {journal} {\bibinfo  {journal} {Phys. Rev. B}\ }\textbf {\bibinfo {volume}
  {54}},\ \bibinfo {pages} {11169} (\bibinfo {year} {1996})}\BibitemShut
  {NoStop}%
\bibitem [{\citenamefont {Kresse}\ and\ \citenamefont
  {Joubert}(1999)}]{Kresse1999}%
  \BibitemOpen
  \bibfield  {author} {\bibinfo {author} {\bibfnamefont {G.}~\bibnamefont
  {Kresse}}\ and\ \bibinfo {author} {\bibfnamefont {D.}~\bibnamefont
  {Joubert}},\ }\href {\doibase 10.1103/PhysRevB.59.1758} {\bibfield  {journal}
  {\bibinfo  {journal} {Phys. Rev. B}\ }\textbf {\bibinfo {volume} {59}},\
  \bibinfo {pages} {1758} (\bibinfo {year} {1999})}\BibitemShut {NoStop}%
\bibitem [{\citenamefont {Monkhorst}\ and\ \citenamefont
  {Pack}(1976)}]{Monkhorst1976}%
  \BibitemOpen
  \bibfield  {author} {\bibinfo {author} {\bibfnamefont {H.~J.}\ \bibnamefont
  {Monkhorst}}\ and\ \bibinfo {author} {\bibfnamefont {J.~D.}\ \bibnamefont
  {Pack}},\ }\href {\doibase 10.1103/PhysRevB.13.5188} {\bibfield  {journal}
  {\bibinfo  {journal} {Phys. Rev. B}\ }\textbf {\bibinfo {volume} {13}},\
  \bibinfo {pages} {5188} (\bibinfo {year} {1976})}\BibitemShut {NoStop}%
\bibitem [{\citenamefont {Perdew}\ and\ \citenamefont {Wang}(1992)}]{Perdew92}%
  \BibitemOpen
  \bibfield  {author} {\bibinfo {author} {\bibfnamefont {J.~P.}\ \bibnamefont
  {Perdew}}\ and\ \bibinfo {author} {\bibfnamefont {Y.}~\bibnamefont {Wang}},\
  }\href {\doibase 10.1103/PhysRevB.46.12947} {\bibfield  {journal} {\bibinfo
  {journal} {Phys. Rev. B}\ }\textbf {\bibinfo {volume} {46}},\ \bibinfo
  {pages} {12947} (\bibinfo {year} {1992})}\BibitemShut {NoStop}%
\bibitem [{\citenamefont {Hanfland}\ \emph {et~al.}(1989)\citenamefont
  {Hanfland}, \citenamefont {Beister},\ and\ \citenamefont
  {Syassen}}]{Hanfland1989}%
  \BibitemOpen
  \bibfield  {author} {\bibinfo {author} {\bibfnamefont {M.}~\bibnamefont
  {Hanfland}}, \bibinfo {author} {\bibfnamefont {H.}~\bibnamefont {Beister}}, \
  and\ \bibinfo {author} {\bibfnamefont {K.}~\bibnamefont {Syassen}},\ }\href
  {\doibase 10.1103/PhysRevB.39.12598} {\bibfield  {journal} {\bibinfo
  {journal} {Phys. Rev. B}\ }\textbf {\bibinfo {volume} {39}},\ \bibinfo
  {pages} {12598} (\bibinfo {year} {1989})}\BibitemShut {NoStop}%
\bibitem [{\citenamefont {Duclaux}\ \emph {et~al.}(1992)\citenamefont
  {Duclaux}, \citenamefont {Nysten}, \citenamefont {Issi},\ and\ \citenamefont
  {Moore}}]{Duclaux1992}%
  \BibitemOpen
  \bibfield  {author} {\bibinfo {author} {\bibfnamefont {L.}~\bibnamefont
  {Duclaux}}, \bibinfo {author} {\bibfnamefont {B.}~\bibnamefont {Nysten}},
  \bibinfo {author} {\bibfnamefont {J.-P.}\ \bibnamefont {Issi}}, \ and\
  \bibinfo {author} {\bibfnamefont {A.~W.}\ \bibnamefont {Moore}},\ }\href
  {\doibase 10.1103/PhysRevB.46.3362} {\bibfield  {journal} {\bibinfo
  {journal} {Phys. Rev. B}\ }\textbf {\bibinfo {volume} {46}},\ \bibinfo
  {pages} {3362} (\bibinfo {year} {1992})}\BibitemShut {NoStop}%
\bibitem [{\citenamefont {Shi}\ \emph {et~al.}(2009)\citenamefont {Shi},
  \citenamefont {Pugno}, \citenamefont {Cheng},\ and\ \citenamefont
  {Gao}}]{Shi2009}%
  \BibitemOpen
  \bibfield  {author} {\bibinfo {author} {\bibfnamefont {X.}~\bibnamefont
  {Shi}}, \bibinfo {author} {\bibfnamefont {N.~M.}\ \bibnamefont {Pugno}},
  \bibinfo {author} {\bibfnamefont {Y.}~\bibnamefont {Cheng}}, \ and\ \bibinfo
  {author} {\bibfnamefont {H.}~\bibnamefont {Gao}},\ }\href {\doibase
  10.1063/1.3253423} {\bibfield  {journal} {\bibinfo  {journal} {Appl. Phys.
  Lett.}\ }\textbf {\bibinfo {volume} {95}},\ \bibinfo {pages} {163113}
  (\bibinfo {year} {2009})}\BibitemShut {NoStop}%
\bibitem [{\citenamefont {Bichoutskaia}\ \emph {et~al.}(2008)\citenamefont
  {Bichoutskaia}, \citenamefont {Popov},\ and\ \citenamefont
  {Lozovik}}]{Bichoutskaia2008}%
  \BibitemOpen
  \bibfield  {author} {\bibinfo {author} {\bibfnamefont {E.}~\bibnamefont
  {Bichoutskaia}}, \bibinfo {author} {\bibfnamefont {A.~M.}\ \bibnamefont
  {Popov}}, \ and\ \bibinfo {author} {\bibfnamefont {Y.~E.}\ \bibnamefont
  {Lozovik}},\ }\href {\doibase https://doi.org/10.1016/S1369-7021(08)70120-2}
  {\bibfield  {journal} {\bibinfo  {journal} {Mater. Today}\ }\textbf {\bibinfo
  {volume} {11}},\ \bibinfo {pages} {38} (\bibinfo {year} {2008})}\BibitemShut
  {NoStop}%
\end{thebibliography}%


%

\end{document}